\documentclass[twocolumn,showpacs,amsmath,amssymb,superscriptaddress]{revtex4-1}
\usepackage{dcolumn}
\usepackage{color}
\usepackage{graphicx}
\usepackage{amsmath}
\usepackage{bm}
\usepackage{url}

\begin{document}

\title{
Spectroscopy of quadrupole and octupole states in rare-earth nuclei from a Gogny force
}

\author{K.~Nomura}
%\email{nomura@ganil.fr}
\affiliation{Grand Acc\'el\'erateur National d'Ions Lourds,
CEA/DSM-CNRS/IN2P3, B.P.~55027, F-14076 Caen Cedex 5, France}

\author{R.~Rodr\'iguez-Guzm\'an}
\affiliation{Physics Department, Kuwait University, 13060 Kuwait, Kuwait}

\author{L.~M.~Robledo}
\affiliation{Departamento de F\'\i sica Te\'orica, Universidad
Aut\'onoma de Madrid, E-28049 Madrid, Spain}

\date{\today}

\begin{abstract}
Collective quadrupole and octupole states are described in a series of 
Sm and Gd isotopes within the framework of the interacting boson model 
(IBM), whose Hamiltonian parameters are deduced from mean field 
calculations with the Gogny energy density functional. The link between 
both frameworks is the ($\beta_2\beta_3$) potential energy surface 
computed within the Hartree-Fock-Bogoliubov framework in the case of 
the Gogny force. The diagonalization of the IBM Hamiltonian provides 
excitation energies and transition strengths of an assorted set of 
states including both positive and negative parity states. The 
resultant spectroscopic properties are compared with the available 
experimental data and also with the results of the configuration mixing 
calculations with the Gogny force within the generator coordinate 
method (GCM). The structure of excited $0^{+}$ states and its 
connection with double octupole phonons is also addressed. The model is 
shown to describe the empirical trend of the low-energy quadrupole and 
octupole collective structure fairly well, and turns out to be 
consistent with GCM results obtained with the Gogny force.  
\end{abstract}



\maketitle

\section{Introduction}

The study of the equilibrium shapes and the corresponding excitation 
spectra of atomic nuclei is one of the recurrent themes in nuclear 
structure physics. Most of the deformed medium-heavy and heavy nuclei 
exhibit  reflection-symmetric ground states.  However, in some regions 
of the nuclear chart, there is an onset of reflection-asymmetric shapes 
driven by specific shell effects. In quadrupole deformed nuclei, a 
characteristic feature  of  octupole deformation is the 
alternating-parity rotational band formed by the even-spin positive 
parity states and alternating  odd-spin negative-parity states, connected with 
each other by enhanced electric dipole transitions \cite{butler96}.

In the framework of  the spherical shell model,  octupolarity arises as 
a result of the coupling between the ($l,j$) orbitals in a major shell 
and the unique-parity ($l+3,j+3$) intruders from the next major shell. 
Within this context, illustrative examples are the rare-earth nuclei 
with the proton number $Z\approx 56$ and the neutron number $N\approx 
88$ as well as the light actinides with $Z\approx 88$ and $N\approx 
134$. In the light actinides case, the coupling of both  neutron (i.e., 
$1g_{9/2}$ and $0j_{15/2}$) and proton (i.e., $1f_{7/2}$ and 
$0i_{13/2}$) single-particle states leads to octupole deformed ground 
states \cite{butler91,butler96}. A recent Coulomb excitation study has 
revealed, for the first time, unambiguous evidences of static octupole 
deformation in $^{224}$Ra \cite{gaffney13}.

In this work, we study the impact of octupole correlations on the  
ground state and the associated low-lying collective spectra of the 
nuclei $^{146-156}$Sm and $^{148-158}$Gd. 
We consider both quadrupole and octupole degrees of 
freedom. The selected nuclei belong to a  region of the nuclear chart 
where octupole correlations are expected to play an important role and 
therefore, represent a valuable testing ground for the considered 
theoretical approximations. Indeed, the experimental observation of  
octupole correlations at medium spin, as well as the crossing of the 
octupole and the ground-state bands, point to the coexistence of 
reflection symmetric and asymmetric structures in both $^{150}$Sm 
\cite{urban87} and $^{148}$Sm \cite{urban91}. From the experimental 
point of view, four low-lying negative-parity bands have already been 
identified in $^{152}$Sm \cite{garrett09}. The emerging pattern of 
excitations, suggests a complex shape coexistence in this nucleus. 
Moreover, the nucleus  $^{152}$Sm has been  identified \cite{casten01} 
as an example of the X(5) critical point symmetry \cite{iachello01}. 
The nature of many low-lying excited $0^+$ states in rare-earth nuclei 
has also attracted much attention. For example, thirteen excited 
$0^{+}$ states have already been identified for $^{158}$Gd 
\cite{lesher02}. Within the $spdf$-IBM framework, many of the observed 
$0^+$ states have been attributed to the coupling of two octupole 
phonons \cite{zamfir02}.

Keeping in mind the experimental findings mentioned above, it is 
interesting and timely to consider a systematic analysis of the 
quadrupole-octupole collectivity in rare-earth nuclei. The breaking of 
reflection symmetry  and the associated low-lying negative-parity  
states have been addressed using various theoretical frameworks: 
self-consistent mean-field  
\cite{marcos83,naza84b,naza85,bonche86,bonche88,egido91,robledo10,robledo11,rayner12,robledo13}, 
algebraic 
\cite{scholten78,engel87,taka88,kusnezov88,cottle98}, 
collective phenomenological 
\cite{bizzeti04,bonatsos05,lenis06,bizzeti08,bizzeti10,jolos12,minkov12,bizzeti13}, 
and cluster
\cite{iachello82,daley86a,shneidman02} models. 
A large number of calculations for nuclei with static and/or dynamical 
octupole deformations have already been reported 
\cite{naza84b,bonche86,bonche88,egido91,robledo10,robledo11,rayner12,robledo13,zhang10,lu12,lu14}. 
In particular, the nuclear energy density functional (EDF) framework, 
both at the mean-field level and beyond, provides a reasonably  
accurate description of the properties of the negative- and 
positive-parity states all  over the nuclear chart \cite{ben03rev}. 
Both non-relativistic \cite{Skyrme,VB,Gogny} and relativistic 
\cite{Vre05,Nik11rev} EDFs have already been  applied in both 
mean-field and  beyond mean-field studies of medium-heavy and heavy 
mass nuclei. The description of the  excitation spectra and transition 
rates requires the inclusion of dynamical (i.e., beyond mean-field) 
correlations associated with the restoration of the broken  symmetries 
and/or fluctuations in the collective parameters (i.e., generating 
coordinates) \cite{rayner02,ben03rev,rayner12,robledo13}. Within this 
context, the projection of the intrinsic (i.e., symmetry-broken)  
states onto good parity ones as well as the corresponding configuration 
mixing, in the spirit of the two-dimensional generator coordinate 
method (GCM) \cite{RS},  have been considered recently for nuclei in 
the rare-earth region using the quadrupole $Q_{20}$ and octupole 
$Q_{30}$ moments as generating coordinates \cite{rayner12}. For recent 
GCM study, based on $Q_{30}$-constrained mean-field states, the reader 
is also referred to Ref.~\cite{robledo12}. 

In this work we first carry out ($Q_{20}, Q_{30}$)-constrained 
Hartree-Fock-Bogoliubov (HFB) calculations based on the Gogny-EDF 
\cite{Gogny}. Such calculations provide us with the corresponding 
(axially symmetric) mean-field potential energy surfaces
(PES). Subsequently, 
in order to obtain the spectrum and  wave functions of the excited states, 
we employ the interacting boson model (IBM) \cite{Nom08}. The essence 
of our method is to determine  the parameters of an appropriate IBM 
Hamiltonian by calculating the associated bosonic PES so that it matches
the  Gogny-HFB  PES. The IBM Hamiltonian  
resulting from our fermion-to-boson mapping procedure is then  used in 
spectroscopic calculations. A similar mapping  has been used in 
previous studies of low-lying quadrupole states 
\cite{Nom10,Nom11rot,Nom12tri} and shape coexistence \cite{Nom12sc}. 
Recently, the method  \cite{Nom08} has been extended to describe 
quadrupole-octupole correlations and  shape transitions in the light 
actinide and rare-earth regions \cite{nom13oct,nom14} based on the 
relativistic  DD-PC1 EDF.

The same Gogny-EDF can be used along with beyond mean field techniques 
to restore the broken reflection symmetry and compute the properties of 
the lowest lying negative parity state. The excitation energy and 
transition strengths, when compared with the IBM numbers, can be used 
as a benchmark to test the consistency of the mapping procedure. 
Therefore, one of the goals of this study  is to assess the 
fermion-to-boson mapping methodology in the description of 
spectroscopic properties in  rare-earth nuclei. We compare the  IBM 
spectra and transition rates with previous Gogny-GCM calculations for 
the same Sm and Gd nuclei \cite{rayner12} as well as with available 
experimental data. Here, we also refer the reader to the  previous IBM 
study based on the relativistic mean-field (RMF) approximation 
\cite{nom14}. We have used the D1M \cite{D1M} parametrization of the 
Gogny-EDF, which was originally designed to better describe nuclear  
masses. It has been shown 
\cite{rayner10odd-1,rayner10odd-2,rayner10odd-3,rayner12,giuliani14} 
that the D1M parameter set essentially retains the same predictive 
power as the standard and thoroughly  tested Gogny-D1S \cite{D1S} one. 
We have also performed a selected set of calculations based on the D1S 
parametrization in order to examine the robustness of our predictions 
with respect to the particular version of the Gogny-EDF employed. 
However, as the corresponding HFB  \cite{rayner12} and IBM results are 
quite similar, in the present paper we will only focus on calculations 
based on the D1M parameter set.

The paper is organized as follows. In 
Sec.~\ref{sec:Theoretical framework}, we briefly outline the HFB-to-IBM mapping procedure. Next, 
in Sec.~\ref{sec:pes}, we discuss the systematics of the 
($\beta_{20},\beta_{30}$) \footnote{We equally use the  multipole 
moment values $Q_{l0}$ and deformation parameters  $\beta_{l}$ to talk about deformation.} 
PESs obtained for the considered nuclei as 
well as the parameters of the IBM Hamiltonian. The results of the 
spectroscopic calculations are discussed in Sec.~\ref{sec:results}. 
First, in Sec.~\ref{sec:level}, we present the systematics of the 
low-energy spectra and the reduced transition probabilities in 
$^{146-156}$Sm and $^{148-158}$Gd. We will  compare with available 
experimental data as well as with results obtained within the Gogny-GCM 
approximation \cite{rayner12}. Next, in Sec.~\ref{sec:spec} we further 
illustrate the predictive power of the mapped IBM model with a detailed 
discussion of the spectroscopic properties for $^{150}$Sm (a soft 
nucleus along the quadrupole and octupole directions) and $^{158}$Gd (a 
strongly quadrupole deformed nucleus). In order to obtain some insight 
into the nature of the excited $0^+$ states in the studied nuclei, 
their systematics is discussed in Sec.~\ref{excited-zeros}. In 
Sec.\ref{sec:corr}, we discuss the IBM correlation energies and compare 
them with Gogny-GCM results. Finally, Sec.~\ref{sec:summary} is devoted 
to some concluding remarks and work perspectives.

%-----------------------------------------------------------------------

\section{Framework\label{sec:Theoretical framework}}

%-----------------------------------------------------------------------

In this section we  briefly outline  the HFB-to-IBM mapping scheme 
\cite{nom14}. Our starting point is a set of axially symmetric 
$(Q_{20},Q_{30})$-constrained Gogny-HFB calculations  \cite{rayner12}. 
They provide us with the corresponding mean-field potential energy 
surfaces (MFPESs) and the HFB states $|\Phi(Q_{20},Q_{30})\rangle$ for 
the nuclei $^{146-156}$Sm and $^{148-158}$Gd. For simplicity, both the 
quadrupole $Q_{20}$ and the octupole $Q_{30}$ moments are then 
translated into the standard $\beta_{2}$ and  $\beta_3$ mean-field 
deformation parameters.

Subsequently, the MFPESs obtained are mapped into their bosonic 
counterparts, i.e., the IBM potential energy surfaces (IBMPESs). This 
procedure allows us to determine the parameters of the IBM Hamiltonian 
used in the spectroscopic calculations.  The IBM Hamiltonian is 
converted into a potential energy surface by means of a set of coherent 
bosonic states and this IBM-PES is what is used to match the 
Gogny-HFB PES \cite{nom14}. Note that the MFPESs correspond to the 
total HFB energies, i.e., neither mass parameters nor zero point 
(rotational and/or vibrational) quantum corrections are included. 
 
The description of the quadrupole and octupole deformations as well as  
the positive- and negative-parity states within the IBM framework 
requires both positive- and negative-parity bosons. Here, one assumes 
that the low-lying positive-parity states are reasonably well described 
by the pairs of valence nucleons associated to the $s$ and $d$ bosons, 
respectively. On the other hand, negative-parity states are assumed to 
be described by the coupling to octupole $f$ bosons \cite{OAI}. 
Therefore, our entire IBM model space comprises the $s$, $d$ and $f$ 
bosons. For simplicity, we do not distinguish between proton and  
neutron bosons. A more complete description of the low-energy 
collective states would require the inclusion of the dipole $p$ boson 
that could be associated to the spurious center-of-mass motion 
\cite{engel87} or to the giant dipole resonance \cite{sugita96}. This, 
however, lies out of the scope of the present paper and is left for 
future work.
  
The  $sdf$ Hamiltonian used is given by
\begin{eqnarray}
 \label{eq:bh}
\hat H=\epsilon_d\hat n_{d}+\epsilon_f\hat
n_f+\kappa_2\hat Q_2\cdot\hat Q_2+\kappa_2^{\prime}\hat
L_d\cdot\hat L_d+\kappa_3\hat Q_3\cdot\hat Q_3, 
\end{eqnarray}
where the first (second) term stands for the number operator for
the $d$ ($f$) bosons with $\epsilon_d$ ($\epsilon_f$) being the single
$d$ ($f$) boson energy relative to the $s$ boson one. 
The third term represents the quadrupole-quadrupole interaction with 
strength $\kappa_2$. The quadrupole operator is given as 
\begin{eqnarray}
 \hat Q_2=s^{\dagger}\tilde d+d^{\dagger}\tilde
  s+\chi_{dd}[d^{\dagger}\times\tilde
  d]^{(2)}+\chi_{ff}[f^{\dagger}\times\tilde f]^{(2)}
\end{eqnarray}
where $\chi_{dd}$ and $\chi_{ff}$ are parameters. The forth term in Eq.~(\ref{eq:bh}) is the 
rotational one relevant for the $sd$ space. In this case, the angular 
momentum operator $\hat L_d$ reads
\begin{eqnarray}
 \hat
L_d=\sqrt{10}[d^{\dagger}\times\tilde d]^{(1)}
\end{eqnarray} 
The last term in Eq.~(\ref{eq:bh}) is the octupole-octupole
interaction with the strength parameter $\kappa_3$. The octupole 
operator takes the form 
\begin{eqnarray}
 \hat Q_3=s^{\dagger}\tilde f+f^{\dagger}\tilde
  s+\chi_{df}[d^{\dagger}\times\tilde
  f+f^{\dagger}\times\tilde
  d]^{(2)}, 
\end{eqnarray}
with $\chi_{df}$ being a parameter. 

Note, that  Eq.~(\ref{eq:bh}) does not represent the most general form 
for the $sdf$ Hamiltonian. The present form has already been used in 
previous phenomenological IBM studies which have confirmed its 
suitability to describe the available experimental data. The 
Hamiltonian $\hat H^{\textnormal{IBM}}$ of Eq.~(\ref{eq:bh}) can be 
derived from a microscopic octupole-octupole interaction between proton 
and neutron bosons by mapping the totally symmetric state in the IBM-2 
space onto the equivalent one in the IBM-1 space \cite{barfield88}. We 
neglect the dipole-dipole interaction term $\hat L_d\cdot\hat L_f$ 
(with $\hat L_f=\sqrt{28}[d^{\dagger}\times\tilde f]^{(1)}$), because 
it has been shown \cite{cottle98} to be of little relevance for 
low-energy states.

The IBMPES is calculated as the expectation value of the Hamiltonian 
Eq.~(\ref{eq:bh}) in the boson condensate state $|\phi\rangle$ \cite{GK}
\begin{eqnarray}
 \label{eq:coherent}
|\phi\rangle=\frac{1}{\sqrt{N_B}}(\lambda^{\dagger})^{N_B}|-\rangle
\quad
{\textnormal{with}}
\quad
\lambda^{\dagger}=s^{\dagger}+\bar\beta_2d_0^{\dagger}+\bar\beta_3f_0^{\dagger}. 
\nonumber \\
\end{eqnarray}
where $N_B(=n_s+n_d+n_f)$ and $|-\rangle$ denote the total number of 
bosons (i.e., half the number of valence nucleons \cite{OAI}) and the 
inert core, respectively. In the present study, the doubly-magic 
nucleus $^{132}$Sn is assumed to be the inert core. Therefore, $N_B$ 
runs from 6 to 12  (7 to 13) in $^{146-156}$Sm ($^{148-158}$Gd). For 
the quadrupole case ($\lambda=2$) the bosonic  $\bar\beta_2$ and 
fermionic $\beta_2$ deformations can be related as 
$\tilde\beta_2=C_2\beta_2$ \cite{GK}, with $C_2$ being a coefficient. 
Here, as in previous works \cite{nom13oct,nom14}, we assume that 
$\tilde\beta_3=C_3\beta_3$, with $C_3$ being an additional coefficient.

In order to reduce the computational effort, it has been customary in 
many of the previous phenomenological IBM calculations to restrict the 
maximum number of $f$ bosons to $n^{max}_f=1$ in the diagonalization of 
the IBM Hamiltonian. However, as shown in the next section, the 
microscopic PESs may exhibit a sizable  ground state octupole 
deformation which requires a larger number of $f$ bosons in our IBM 
calculations. Therefore both positive- and negative-parity bosons are 
treated on an equal footing. As a consequence, a truncation on 
$n^{max}_f$ is not used and the number of $f$ bosons can  run from 0 to 
$N_B$. This also holds true for the $s$ and $d$ bosons. Let us also 
mention, that previous phenomenological studies (e.g., 
\cite{zamfir01,babilon05}) have also suggested the need of more 
negative-parity bosons for a better description of the experimental 
data. 

The analytic IBMPES reads 
\begin{eqnarray}
\label{eq:pes}
 E(\bar\beta_{2},
  \bar\beta_{3})
&=&
\frac{N_{B}}{1+\bar\beta_{2}^{2}+\bar\beta_{3}^{2}}
\Big(
\epsilon_{s}^{\prime}+
\epsilon_{d}^{\prime}\bar\beta_{2}^{2}+\epsilon_{f}^{\prime}\bar\beta_{3}^{2}
\Big)
\nonumber \\
&&
+\frac{N_{B}(N_{B}-1)}{(1+\bar\beta_{2}^{2}+\bar\beta_{3}^{2})^2}\times
\nonumber \\
&&\Big[
\kappa_{2}\Big(
2\bar\beta_{2}-\sqrt{\frac{2}{7}}\chi_{dd}\bar\beta_{2}^{2}-\frac{2}{\sqrt{21}}\chi_{ff}\bar\beta_{3}^{2}
\Big)^{2} \nonumber \\
&&-4\kappa_{3}
\Big(\bar\beta_{3}-\frac{2}{\sqrt{15}}\chi_{df}\bar\beta_{2}\bar\beta_{3}
\Big)^{2}
\Big],
\end{eqnarray}
with
\begin{eqnarray}
\label{eq:eps-prime}
&&\epsilon_{s}^{\prime}=5\kappa_{2}-7\kappa_3,\quad
 \epsilon_{d}^{\prime}=\epsilon_{d}+6\kappa_2^{\prime}+(1+\chi_{dd}^2)\kappa_{2}
-\frac{7}{5}\chi_{df}^2\kappa_{3}
\nonumber \\
&&{\textnormal{and}}
\quad
\epsilon_{f}^{\prime}=\epsilon_{f}-\frac{5}{7}\chi_{ff}^{2}\kappa_{2}+(1+\chi_{df}^2)\kappa_{3}.
\end{eqnarray}

The IBMPES $E(\bar\beta_{2},\bar\beta_{3})$  is specified by the 
parameters of the Hamiltonian in Eq.~(\ref{eq:bh}) 
%$\epsilon_d$, $\epsilon_f$, $\kappa_2$, $\chi_{dd}$, 
%$\chi_{ff}$ $\kappa_3$ and $\chi_{df}$ 
plus the coefficients $C_{2}$ 
and $C_{3}$. We have determined those parameters by fitting the IBMPESs 
to the Gogny-D1M MFPESs using the same procedure as in 
Ref.~\cite{Nom10}. Let us remark that, even though a simplified 
Hamiltonian Eq.~(\ref{eq:bh}) is considered, there is still a larger 
number of parameters to be determined, as compared to the $sd$ IBM 
system. Therefore, rather than trying to fit all the parameters at 
once, we first determine the ones relevant for the $sd$ space 
($\epsilon_d$, $\kappa_2$, $\chi_{dd}$, $C_{2}$ and 
$\kappa^{\prime}_2$) and then those associated to the $f$ space as well 
as the ones associated  with the coupling between the two spaces 
($\epsilon_f$, $\kappa_3$, $\chi_{ff}$, $\chi_{df}$ and $C_3$). The 
$\hat L_d\cdot\hat L_d$ term in Eq.~(\ref{eq:bh}) does not contribute 
to the PESs, and therefore its strength $\kappa^{\prime}_2$ is 
determined independently by comparing the fermionic and bosonic 
cranking moment of inertia (see Ref.~\cite{Nom11rot} for details). The 
(fermionic) Thouless-Valatin  \cite{TV} moment of inertia for the 
$2^+_1$ state reads 
\begin{eqnarray}
\label{eq:tv}
 {\cal I}_{\textnormal{TV}}=3/E_{\gamma}. 
\end{eqnarray}
where $E_{\gamma}$ stands for the $2^{+}_{1}$ excitation energy 
obtained from the self-consistent cranking calculation with the 
constraint $\langle\hat J_{x}\rangle=\sqrt{J(J+1)}$, where $\hat J_x$ 
represents the $x$ component of the angular momentum operator. On the 
other hand, the IBM moment of inertia is computed using the coherent 
state $|\phi(\beta,\gamma)\rangle$ and the Schaaser-Brink 
\cite{Schaaser86} expression
\begin{eqnarray}
\label{eq:bmom}
 {\cal I}_{\textnormal{IBM}}=\lim_{\omega\rightarrow\infty}\frac{1}{\omega}\frac{\langle\phi(\beta,\gamma)|\hat
  L_{x}|\phi(\beta,\gamma)\rangle}{\langle\phi(\beta,\gamma)|\phi(\beta,\gamma)\rangle}, 
\end{eqnarray}
with $\omega$ being the cranking frequency. 

Having the parameters 
$\epsilon^{\prime}_d(=\epsilon_d-6\kappa^{\prime}_2)$, $\kappa_2$, 
$\chi_{dd}$ and $C_2$ already determined from the fit of the IBMPES to 
the MFPES in the $sd$ space, the IBM moment of inertia in 
Eq.~(\ref{eq:bmom}) depends only in the  parameter $\kappa^{\prime}_2$ 
whose value is determined so that ${\cal I}_{\textnormal{IBM}}$ is 
equal to the ${\cal I}_{\textnormal{TV}}$ value at  the energy minimum. 

From the diagonalization of the $sdf$-IBM Hamiltonian, we have 
obtained both the energies and wave functions of the spectrum which are labeled by 
total spin and parity quantum numbers. We have used the computer 
program OCTUPOLE \cite{OCTUPOLE}. The reduced electromagnetic 
transition probabilities $B(E\lambda;J\rightarrow J^{\prime})=|\langle 
J^{\prime}||\hat T^{(E\lambda)}||J\rangle|^2/(2J+1)$ ($\lambda=1,2,3$) 
are  then computed using the resulting IBM wave functions. Here, $J$ 
($J^{\prime}$) denotes the spin for the initial (final) state. Of 
particular interest for the present study are the dipole 
E1, quadrupole E2, and octupole E3 transition probabilities defined in terms of the 
operators 
\begin{eqnarray}
&&T^{(E1)}=e_1[d^{\dagger}\times\tilde f+d^{\dagger}\times\tilde
 f]^{(1)} \\
&&T^{(E2)}=e_2\hat Q_2 \\
&&T^{(E3)}=e_3\hat Q_3
\end{eqnarray}
where $\hat Q_2$ and $\hat Q_3$ are the  quadrupole and
octupole operators appearing in the IBM Hamiltonian and
$e_\lambda$'s are  boson effective charges which are 
kept constant for all the  considered nuclei. 
Their values are taken from previous phenomenological IBM
studies ($e_1=0.01$ $e$b$^{1/2}$ \cite{babilon05}, $e_2=0.13$ $e$b 
\cite{babilon05} and $e_3=0.099$ $e$b$^{3/2}$ \cite{taka88}). It has been 
shown that they provide a reasonable overall description of the experimental 
data. However, they are not the ones derived microscopically.  Therefore, 
in the following discussions, one should always keep in mind that there 
is some extra freedom in the overall scale of the calculated IBM transitions. 

% ----------------------------------------------------------------------

\section{Mean-field potential energy surfaces and the parameters of the 
IBM Hamiltonian \label{sec:pes}}

% ----------------------------------------------------------------------
 
In this section, we discuss the systematics of the MFPESs and IBMPESs 
as well as the parameters of the IBM Hamiltonian obtained along the 
lines described in Sec. \ref{sec:Theoretical framework}.

\begin{figure*}[ctb!]
\begin{center}
\includegraphics[width=\linewidth]{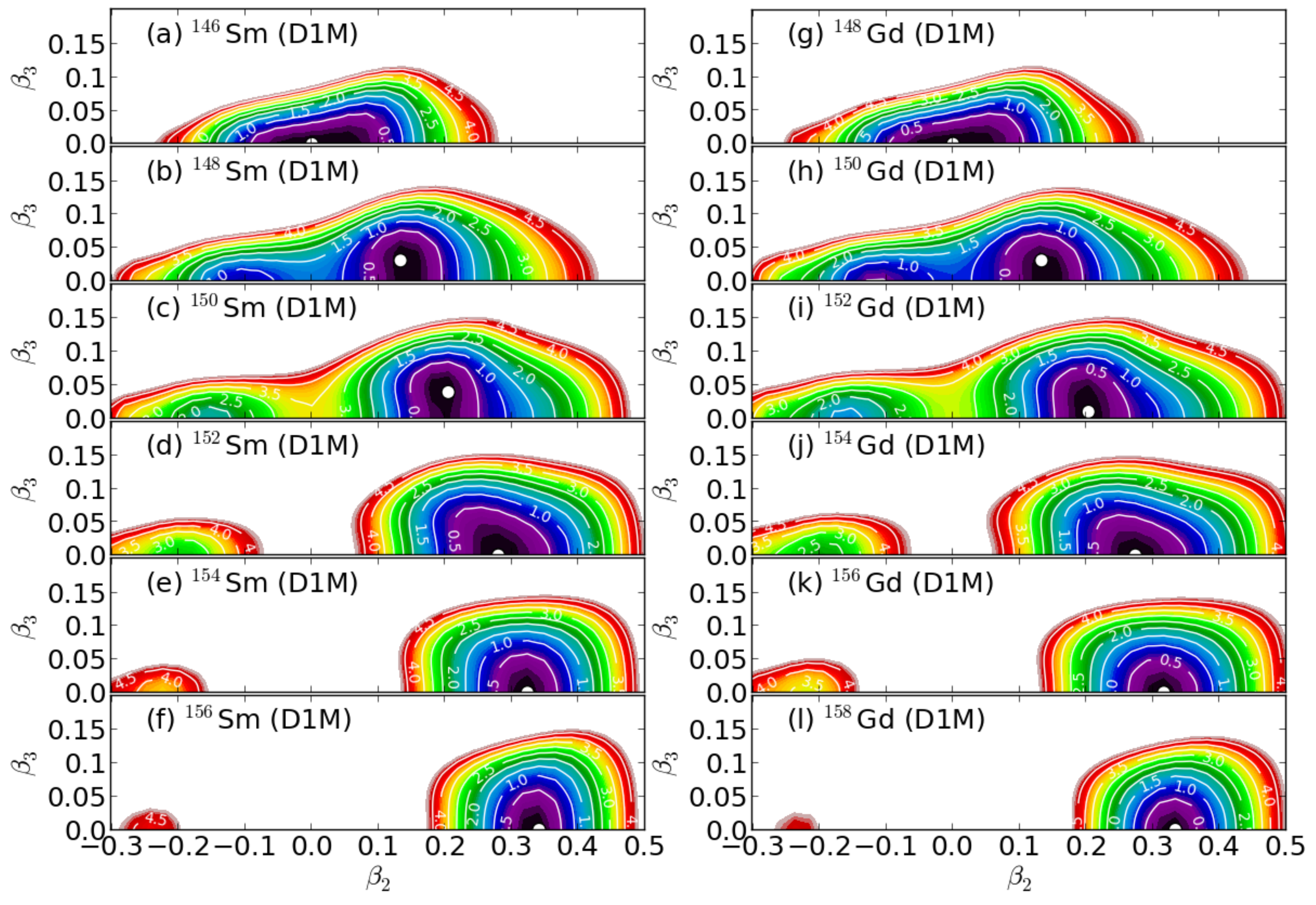}
\caption{(Color online) Axially symmetric ($\beta_2$, $\beta_3$)
 potential energy surfaces for the nuclei $^{146-156}$Sm and $^{148-158}$Gd 
 calculated within the constrained Gogny-HFB approach based on the 
 D1M parametrization. The contour lines join points with the same energy 
 (in MeV) and the color scale varies in steps of 100 keV. The energy 
 difference between neighboring contours is 0.5 MeV. These ($\beta_2, \beta_3$) 
 energy surfaces are symmetric with respect to the $\beta_3=0$ axis. Thus, 
 they are only plotted  for $\beta_3\geqslant 0$. For each nucleus the absolute
 minimum is identified by an open circle.}
\label{fig:hfb_pes}
\end{center}
\end{figure*}
%\marginpar{The open circle of the minima is not visible}

\begin{figure*}[ctb!]
\begin{center}
\includegraphics[width=\linewidth]{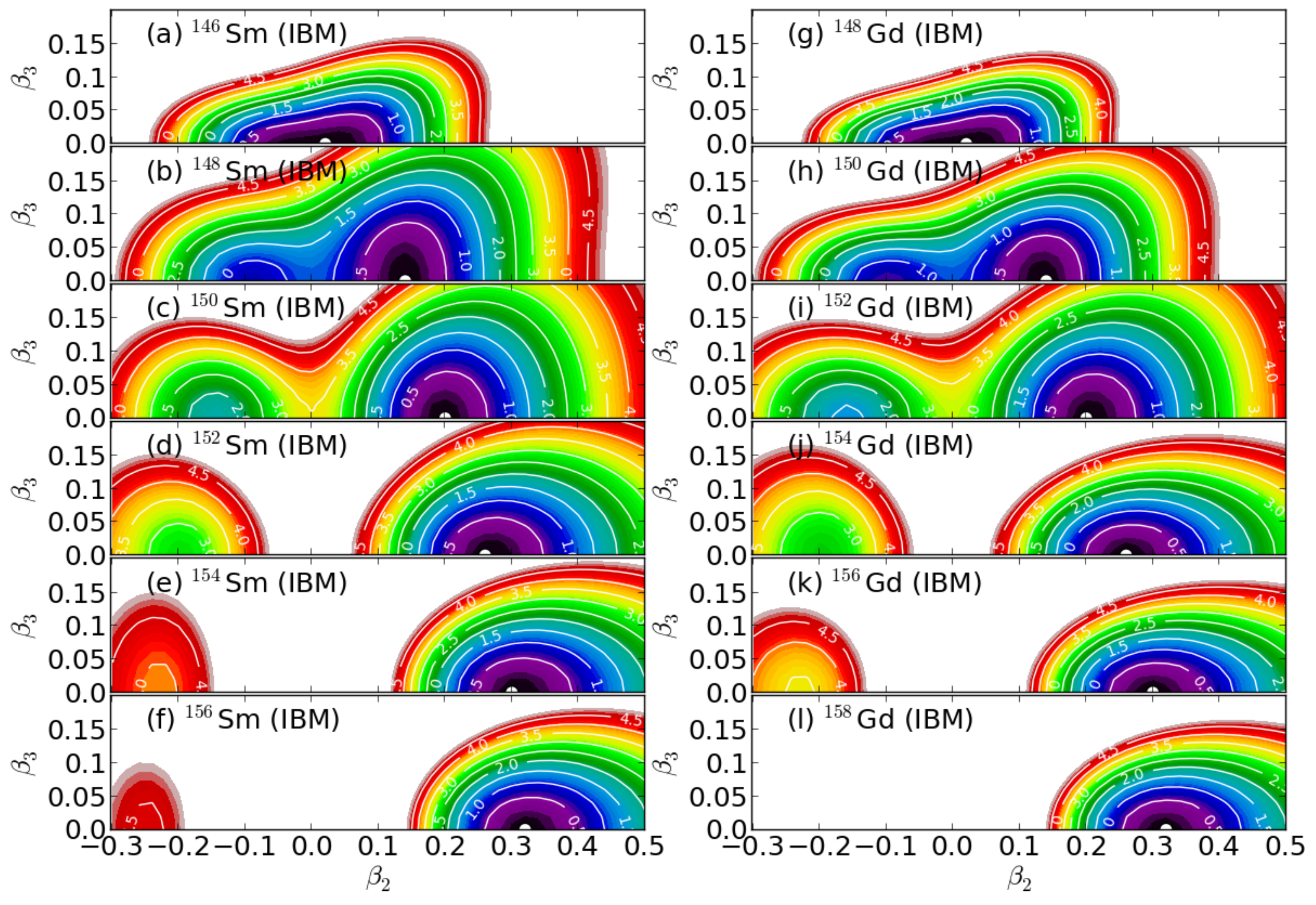}
\caption{(Color online) The same as  Fig.~\ref{fig:hfb_pes}
 but for the mapped IBM potential energy surfaces.}
\label{fig:mapped_pes}
\end{center}
\end{figure*}

\begin{figure*}[ctb!]
\begin{center}
\includegraphics[width=0.8\linewidth]{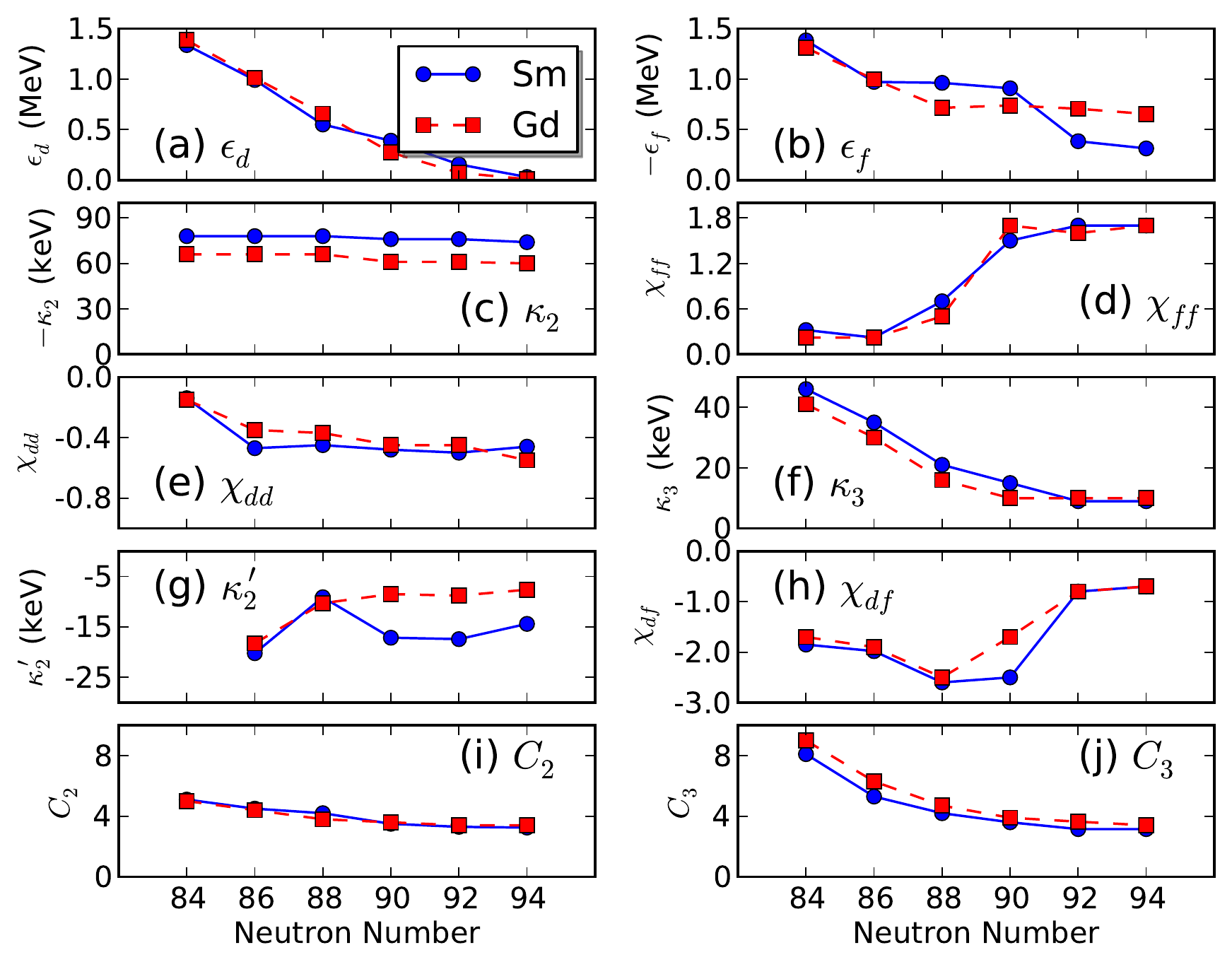}
\caption{(Color online) The parameters of the  $sdf$ IBM Hamiltonian
 $\hat H$ in 
 Eq.~(\ref{eq:bh}), as well as the proportionality coefficients 
 $C_2$ and $C_3$, are plotted as functions of the neutron number for
 the considered nuclei. The parameters $\chi_{dd}$,
 $\chi_{ff}$, $\chi_{df}$, $C_2$ and $C_3$ are are dimensionless. }
\label{fig:para}
\end{center}
\end{figure*}

The axially symmetric Gogny-D1M MFPESs are shown in 
Fig.~\ref{fig:hfb_pes} for $^{146-156}$Sm and $^{148-158}$Gd. 
The MFPESs of some of the Sm isotopes have already been presented in 
Ref.~\cite{rayner12} as illustrative examples. However, for the sake of 
completeness, in the figure we have included all the MFPESs both for Sm and 
Gd nuclei. For the sake of presentation, the plots in the figure 
correspond to $-0.3\leqslant\beta_{2}\leqslant 0.5$ and 
$0.0\leqslant\beta_3\leqslant 0.2$ as well as to an energy range of 5 
MeV from the absolute minimum. We have tested, that the previous ranges 
are enough to  describe the considered low-energy collective states and 
used them to build our IBM Hamiltonian.

A spherical reflection-symmetric ground state is predicted for the 
nuclei $^{146}$Sm [panel (a)] and $^{148}$Gd [panel (g)], respectively. 
On the other hand, the MFPESs become soft for isotopes with 
neutron numbers $N=86$ and $88$, indicating that the Gogny-HFB 
approximation can only be considered as a valuable starting point in 
such nuclei but beyond mean-field correlations should be taken into 
account  \cite{rayner12}. Moreover, the $N=88$ isotopes  exhibit the 
softest MFPESs with a shallow minimum at a non-zero $\beta_3$ value. 
One also sees that the MFPESs become steeper along the $\beta_3$ 
direction  for  isotopes with $N\geqslant 90$. Similar trends have been 
found up to $N=88$ in previous RMF calculations 
\cite{zhang10,nom14}, based on the EDFs PK1 \cite{long04} and DD-PC1 
\cite{DDPC1}, respectively. However in those calculations, the octupole 
minima are more pronounced than ours. In fact, the previous study with 
the relativistic functional DD-PC1 \cite{nom14} suggested that the 
potential energy surface is much more softer along $\beta_3$ direction. 
The same trend was found for isotopes with $N\geqslant 90$. 

As already discussed in Ref.~\cite{rayner12}, there is no essential 
difference between the overall topology of the MFPESs obtained with the 
Gogny-D1M and Gogny-D1S EDFs. However, at a quantitative level, the 
latter provides MFPESs with slightly deeper absolute  minima than the 
former. Nevertheless, such a difference turns out to be too small to 
significantly affect neither the IBM parameters nor the  energies and  
wave functions of the excited states. With this in mind, in what 
follows only results based on the Gogny-D1M EDF will be discussed.
 
In Fig.~\ref{fig:mapped_pes} we have depicted the (mapped) IBMPESs. 
First, we observe that they  are much flatter than the HFB MFPESs (see, 
Fig.~\ref{fig:hfb_pes}). This is a common feature of the IBM framework 
already found in previous studies \cite{Nom08,Nom10}. The reason is 
that IBM's model space is rather limited and only comprises pairs of 
valence nucleons. This  leads to flat IBMPESs for larger deformations. 
However, one should keep in mind that within the considered 
fermion-to-boson mapping, the topology far away from the absolute 
minimum is not relevant as long as we restrict our analysis to the 
low-lying collective states. Hence, we only focus on reproducing the 
curvatures of the Gogny-D1M MFPESs in the neighborhood (a 5 MeV window) 
of the absolute minimum, along both the $\beta_2$ and $\beta_3$ 
directions. 

Second, we note that, for $N=86$ and 88 isotopes, the MFPES predicts a 
shallow absolute minimum at non-zero $\beta_3$ values 
[Fig.~\ref{fig:hfb_pes}] while in the corresponding IBMPES the absolute 
minimum is found at $\beta_3=0$ [Fig.~\ref{fig:mapped_pes}]. However, 
as the depth of this absolute minimum in the MFPESs differs by at most 
tens of keVs  from the saddle point on the $\beta_3=0$ axis, we assume 
that the discrepancy of the absolute minimum point, that is not deep 
enough in energy, between the MFPES and the IBMPES is not of crucial 
importance for the final result. 

Bearing those in mind, the  IBMPESs in Fig.~\ref{fig:mapped_pes} 
closely follow, for each of the considered nuclei, the basic topology 
as well as the overall systematic trend of the Gogny-HFB ones shown in 
Fig.~\ref{fig:hfb_pes}. 

In Fig.~\ref{fig:para}, the IBM parameters for the considered Sm and Gd 
nuclei are plotted as  functions of neutron number. As can be observed 
in panels (a) and (b), the single $d$ ($\epsilon_d$) and $f$ 
($\epsilon_f$) boson energies decrease as functions of neutron number. 
%The origin of such a behavior is still unclear and requires further
%study. 
%Their boson-number dependence is explained if one starts from the most
%general form of the $sdf$ Hamiltonian. 
%The most general form of the $sdf$ IBM Hamiltonian contains several
%two-body terms, which are reduced to kinetic energies of $d$ and $f$
%bosons ($d^{\dagger}\cdot\tilde d$ and $f^{\dagger}\cdot\tilde f$, respectively) multiplied with
%boson-number dependent factors. 
%These terms are absorbed in $\epsilon_d$ and $\epsilon_f$, thereby 
%making the parameters boson-number dependent \cite{IBM}. 
%Therefore, the behaviors of the derived $\epsilon_d$ [panel (a)] and
%$\epsilon_f$ [panel (b)] could reflect these two-body terms in the 
%general IBM Hamiltonian, that should be in principle included but are
%neglected in the present Hamiltonian in Eq.~(\ref{eq:bh}). 
From a microscopic point of view, as already discussed in the
context of the $sd$ IBM-2 \cite{OAI,taka81,taka85} model, the decrease
of $\epsilon_d$ could be related to the
coupling of the {\it unperturbed} $d$ boson with other types of bosons not yet 
explicitly included in the model space. 
%Let us try to understand the boson-number dependence of $\epsilon_d$ and
%$\epsilon_f$ from a simple perspective. 
%There is an alternative explanation: 
%The most general form of the $sdf$ IBM Hamiltonian contains several
%two-body terms, which are reduced to kinetic energies of $d$ and $f$
%bosons ($d^{\dagger}\cdot\tilde d$ and $f^{\dagger}\cdot\tilde f$, respectively) multiplied with
%boson-number dependent factors. 
%When one derives the present Hamiltonian in Eq.~(\ref{eq:bh}) from a general IBM Hamiltonian, these
%terms are absorbed in $\epsilon_d$ and $\epsilon_f$, thereby  
%making the parameters boson-number dependent \cite{IBM}. 
Alternatively, when one derives the form of the IBM Hamiltonian in
Eq.~(\ref{eq:bh}) from a general 
$sdf$ IBM Hamiltonian, several two-body terms of the general IBM
Hamiltonian, that are reduced to the kinetic energies of $d$ and $f$
bosons multiplied with the boson-number dependent
factors, are absorbed in $\epsilon_d$ and $\epsilon_f$, thereby  
making the parameters vary significantly with boson number \cite{IBM}. 

%An alternative explanation is that, in the IBM Hamiltonian, there are those two-body terms which are reduced
%to the kinetic energies of $d$ and $f$ bosons 
%multiplied with boson-number dependent factors and that these terms are
%absorbed in $\epsilon_d$ and $\epsilon_f$, thereby  
%making the parameters boson-number dependent \cite{IBM}. 

The coupling strength of the  quadrupole-quadrupole interaction 
$\kappa_2$, shown in panel (c), is almost constant. A similar trend has 
been found in the IBM study based on the RMF approximation 
\cite{nom14}. A sudden change is observed in the parameter $\chi_{ff}$, 
plotted  in panel (d), around  $N=88$ and is correlated with the 
significant change observed in the MFPESs (see, 
Fig.~\ref{fig:hfb_pes}). On the other hand, at variance with our 
previous $sd$ IBM study in the same mass region \cite{Nom10}, the 
parameter $\chi_{dd}$ [panel (e)] is rather constant. Compared to the 
quadrupole-quadrupole coupling  $\kappa_{2}$ [panel (c)], the strength 
of the octupole-octupole interaction $\kappa_3$ [panel (f)] exhibits a 
gradual decrease with increasing neutron number. 

In panel (g) of the same figure, we have plotted the strength 
$\kappa_2^{\prime}$ of the $\hat L_d\cdot\hat L_d$ term 
Eq.~(\ref{eq:bh}). Its negative value, for all the studied nuclei, 
leads to the lowering of the  positive-parity yrast states 
\cite{Nom11rot}. Note that $\kappa_2^{\prime}$ is not considered for 
the spherical nuclei $^{146}$Sm and $^{148}$Gd. As shown below, the 
experimental spectra for these nuclei do not exhibit a rotational-like 
structure and, therefore, there is no obvious reason for introducing 
the $\hat L_d\cdot\hat L_d$ term in the corresponding calculations. The 
parameters $\chi_{df}$ [panel (h)] exhibits a pronounced isotopic 
dependence with a maximum around $N=88-90$ which correlates well 
with the octupole softness of the MFPESs around the same neutron 
numbers. Both the $C_2$ [panel (i)] and $C_3$ [panel (j)] coefficients 
change smoothly with neutron number \cite{nom14}.

% ----------------------------------------------------------------------

\section{Spectroscopic calculations \label{sec:results}}

% ----------------------------------------------------------------------

In this section, we discuss the results of the calculations with 
the IBM Hamiltonian for 
$^{146-156}$Sm and $^{148-158}$Gd. First, in Sec.~\ref{sec:level}, 
the systematics of the low-energy spectra and the reduced 
transition probabilities in $^{146-156}$Sm and $^{148-158}$Gd is addressed. Next, in 
Sec.~\ref{sec:spec}, the 
spectroscopic properties predicted for the nuclei $^{150}$Sm and 
$^{158}$Gd are discussed in detail. The systematics of the excited $0^+$ states is presented in 
Sec.~\ref{excited-zeros}. Finally, in Sec.\ref{sec:corr}, 
ground state correlation energies are discussed.

% ......................................................................
\subsection{Systematics of the  low-energy spectra and the reduced 
transition probabilities in $^{146-156}$Sm and $^{148-158}$Gd}
\label{sec:level}
%.......................................................................

\begin{figure*}[ctb!]
\begin{center}
\includegraphics[width=0.6\linewidth]{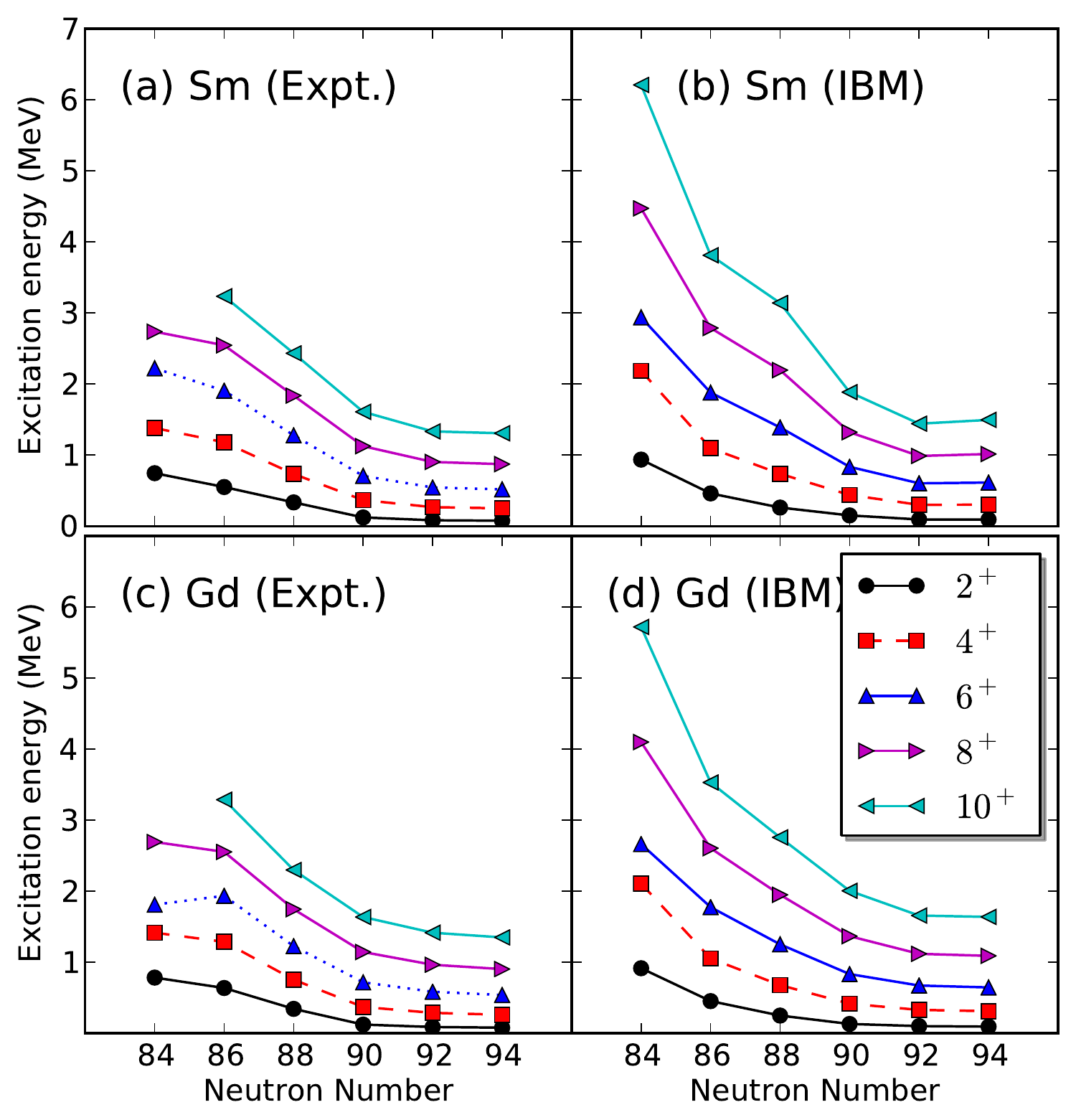}
\caption{(Color online) 
The energy spectra of the lowest-lying even-spin positive-parity states
 up to $J^{\pi}=10^+$ for the considered Sm and
 Gd isotopes. All the experimental data are taken from the NNDC compilation
 \cite{data}. 
}
\label{fig:pos}
\end{center}
\end{figure*}

\begin{figure*}[ctb!]
\begin{center}
\includegraphics[width=0.6\linewidth]{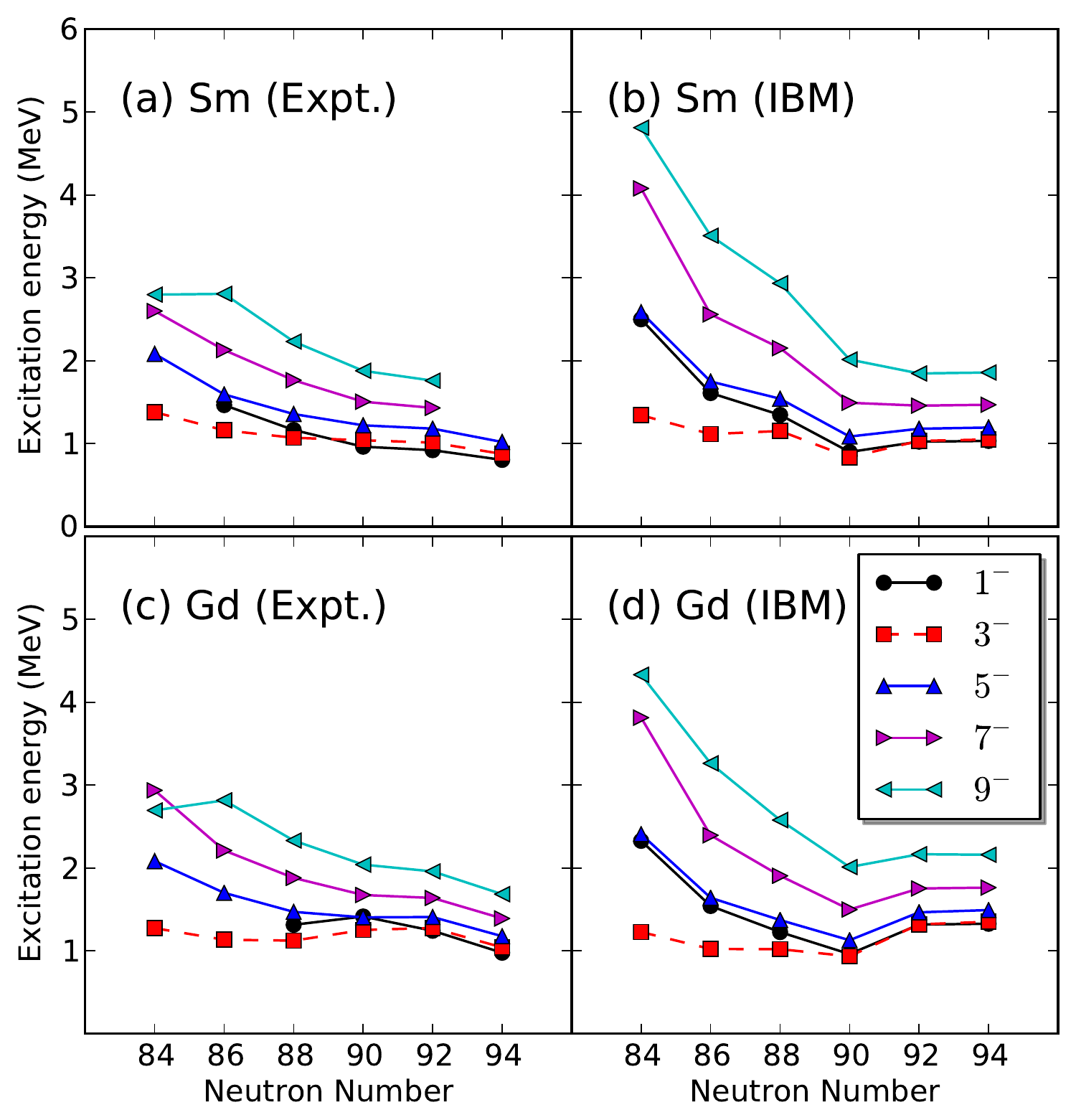}
\caption{(Color online) 
The same as in Fig.~\ref{fig:pos}, but for the lowest-lying
 odd-spin negative-parity states up to $J^{\pi}=9^-$. 
}
\label{fig:neg}
\end{center}
\end{figure*}

In Figs.~\ref{fig:pos} and \ref{fig:neg}  the low-energy positive- and 
negative-parity yrast states, as calculated with the mapped $sdf$ IBM 
Hamiltonian are plotted for the nuclei $^{146-156}$Sm and 
$^{148-158}$Gd. The theoretical results are compared with the available 
experimental data taken from the NNDC compilation \cite{data}.  Since 
our predictions  for  Sm [panels (a) and (b)] and Gd [panels (c) and 
(d)] isotopes are rather similar, we mainly discuss the former.

The lowering of the energies with increasing neutron number $N$ is 
consistent with a shape transition (see, Fig.~\ref{fig:hfb_pes}) to a 
strongly quadrupole deformed configurations. Indeed, the ratios 
$R_{4/2}\equiv E(4^+_1)/E(2^+_1)$=2.33 and 2.38 obtained for 
$^{146,148}$Sm are both close to the vibrational limit while the 
theoretical (experimental) $R_{4/2}$ values for the transitional nuclei 
$^{150,152}$Sm are 2.82 (2.31) and 2.91 (3.01), respectively. Our 
calculations predict a more pronounced rotational character for 
$^{150}$Sm than expected from the  experiment. On the other hand, it is 
remarkable that the $R_{4/2}$ value  for the $^{152}$Sm is exactly the 
same as the X(5) one \cite{iachello01}. For the heavier isotopes, our 
IBM calculations predict well developed rotational bands. For example, 
in the case of $^{154,156}$Sm, we have obtained the ratios 
$R_{4/2}$=3.21 and 3.25, respectively. The theoretical results agree 
reasonably well with the experimental ones except  for the lightest 
isotopes where the energies of the higher spin states are 
overestimated. The reason for the overestimation 
could be the too restricted model space and/or 
Hamiltonian of the IBM that is not rich enough as to
reproduce the peculiar topology of the Gogny-EDF MFPES for the lightest
isotopes. 
We recall that the $\hat L_d\cdot\hat L_d$
term is not included in $^{146}$Sm and $^{148}$Gd as it is of little
importance for these spherical nuclei \cite{Nom11rot}. 
One could introduce this term phenomenologically to fix the 
overestimation, which is however out of scope of the present work. 
%Thus the overestimation for the lightest isotopes could be resolved by
%introducing a priori the $\hat L_d\cdtot\hat L_d$ term. 

The $J^{\pi}=1^-,\ldots,9^-$ states, plotted in Fig.~\ref{fig:neg}, 
display features characteristic of the octupole collectivity. Exception 
made of the $3^-$ states, their excitation energies decrease sharply 
for $84 \le N \le 90$. At variance with the experimental data, the 
theoretical excitation energies increase for $N > 90$ which correlates 
well with the diminishing of the octupole minimum depth observed in the 
MFPESs (see, Fig.~\ref{fig:hfb_pes}). In both isotopic chains, the 
$3^-_1$ state is lower in energy than the $1^-_1$ one. We have also 
found a near degeneracy for the $1^-_1$ and the $5^-_1$ states for 
$N\leqslant 88-90$. This  octupole vibrational feature becomes more 
apparent  for the lighter isotopes. 

\begin{figure*}[ctb!]
\begin{center}
\includegraphics[width=0.6\linewidth]{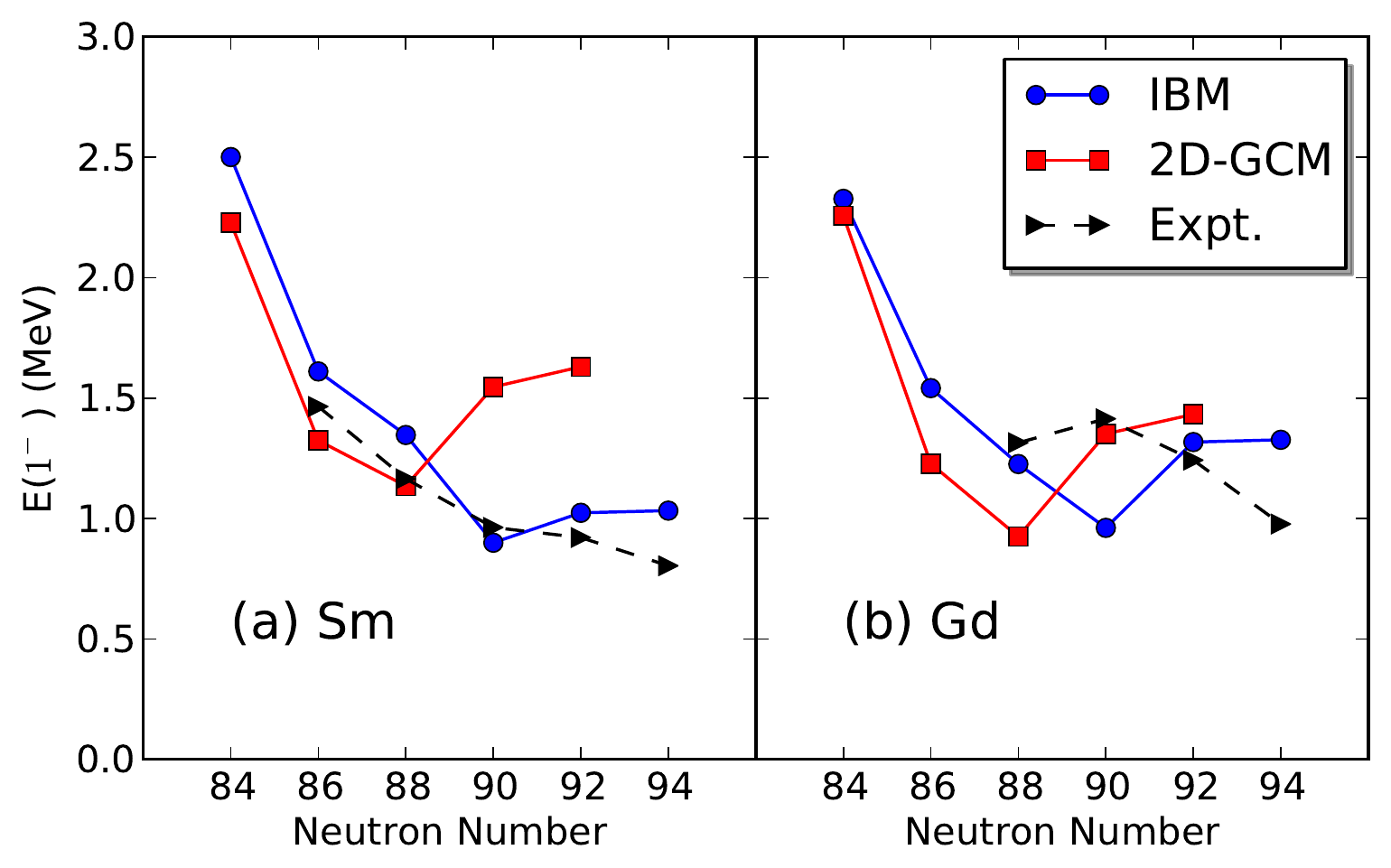}
\caption{(Color online)
The excitation energies of the $1^-_1$ states predicted with the mapped 
IBM Hamiltonian are compared with the ones obtained in the framework of 
two-dimensional GCM calculations \cite{rayner12} for Sm [panel (a)] and 
Gd [panel (b)] nuclei. In both methods, the Gogny-D1M parametrization 
has been used. The experimental energy levels are also included in the 
figure.
}
\label{fig:ibmgcm}
\end{center}
\end{figure*}

In Fig.~\ref{fig:ibmgcm}, we have compared the excitation energies of 
the lowest $1^-_1$ states with the ones obtained in the framework of a 
two-dimensional GCM calculations \cite{rayner12} also with the 
Gogny-D1M EDF. The predicted IBM and GCM values are quite similar for  
$84 \le N \le 88$. In the case of the Sm isotopes both the GCM and IBM 
excitation energies increase with increasing neutron number though the 
former exhibit a more pronounced change than the latter. Similar 
results are obtained for Gd isotopes, exception made of the fact that 
the smallest $1^-$ excitation energy is found  at $N=88$ ($N=90$) in 
the GCM (IBM) calculations.

\begin{figure*}[ctb!]
\begin{center}
\includegraphics[width=0.6\linewidth]{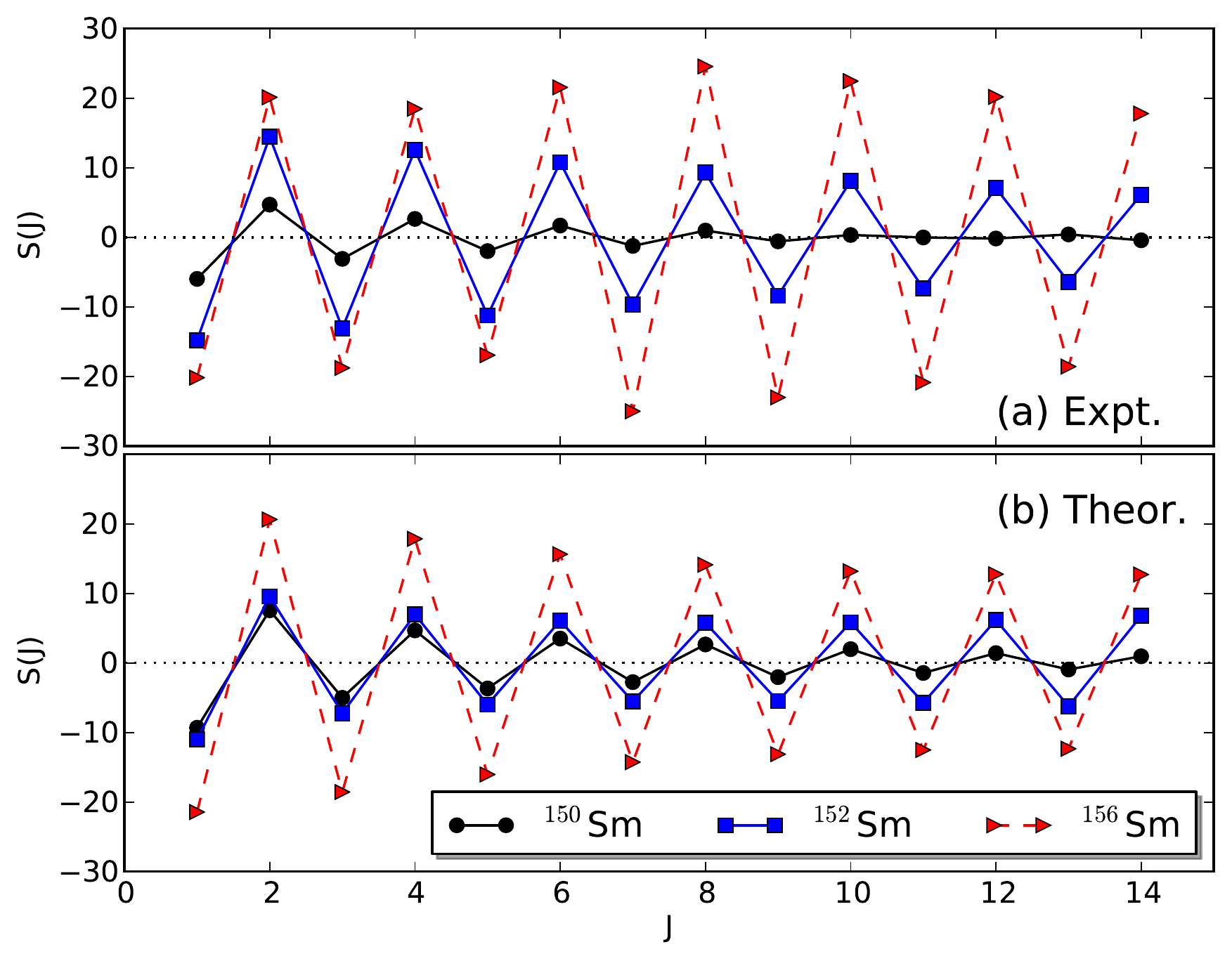}
\caption{(Color online) 
Signature splitting $S(J)$ of $^{150,152,156}$Sm nuclei as a function of
spin $J$. For more details, see the main text.
}
\label{fig:apb}
\end{center}
\end{figure*}

We have studied the quantity
\begin{eqnarray} \label{ref-formula-Rayner}
 S(J)=\frac{[E(J+1)-E(J)]-[E(J)-E(J-1)]}{E(2^+_1)}, 
\end{eqnarray}
which is sensitive to the splitting between the  positive- and 
negative-parity members of a rotational band. In 
Eq.~(\ref{ref-formula-Rayner}), $E(J)$ stands for the excitation energy 
of the $J=0^+,1^-,2^+,3^-,\ldots$ state. Note that, for an ideal 
alternating-parity band, we would obtain an equal energy splitting 
between the positive- and negative-parity states differing by $\Delta 
J=1$. This, in turn, would  lead to $S(J)\approx 0$. On the other hand, 
a non-zero $S(J)$ value indicates a deviation from a pure 
alternating-parity band.
 
In Fig.~\ref{fig:apb} we have plotted $S(J)$, as a function of the spin 
$J$, for $^{150,152,156}$Sm which are taken as representative examples. 
The experimental  data for $^{150}$Sm [panel (a)]  oscillate with $J$ 
but become zero around $J\approx 8^+$. Though larger deviations are 
observed in our calculations [panel (b)] their global trend resembles 
the experimental one. Both theoretically and experimentally, the 
deviation from $S(J)=0$ in $^{156}$Sm is more pronounced than for 
$^{150,152}$Sm. This suggests a deviation from the ideal 
alternating-parity band behavior, and also correlates well with the 
behavior of the Gogny-D1M MFPESs (see, Fig.~\ref{fig:hfb_pes}). 

\begin{figure*}[ctb!]
\begin{center}
\begin{tabular}{cc}
\includegraphics[width=0.6\linewidth]{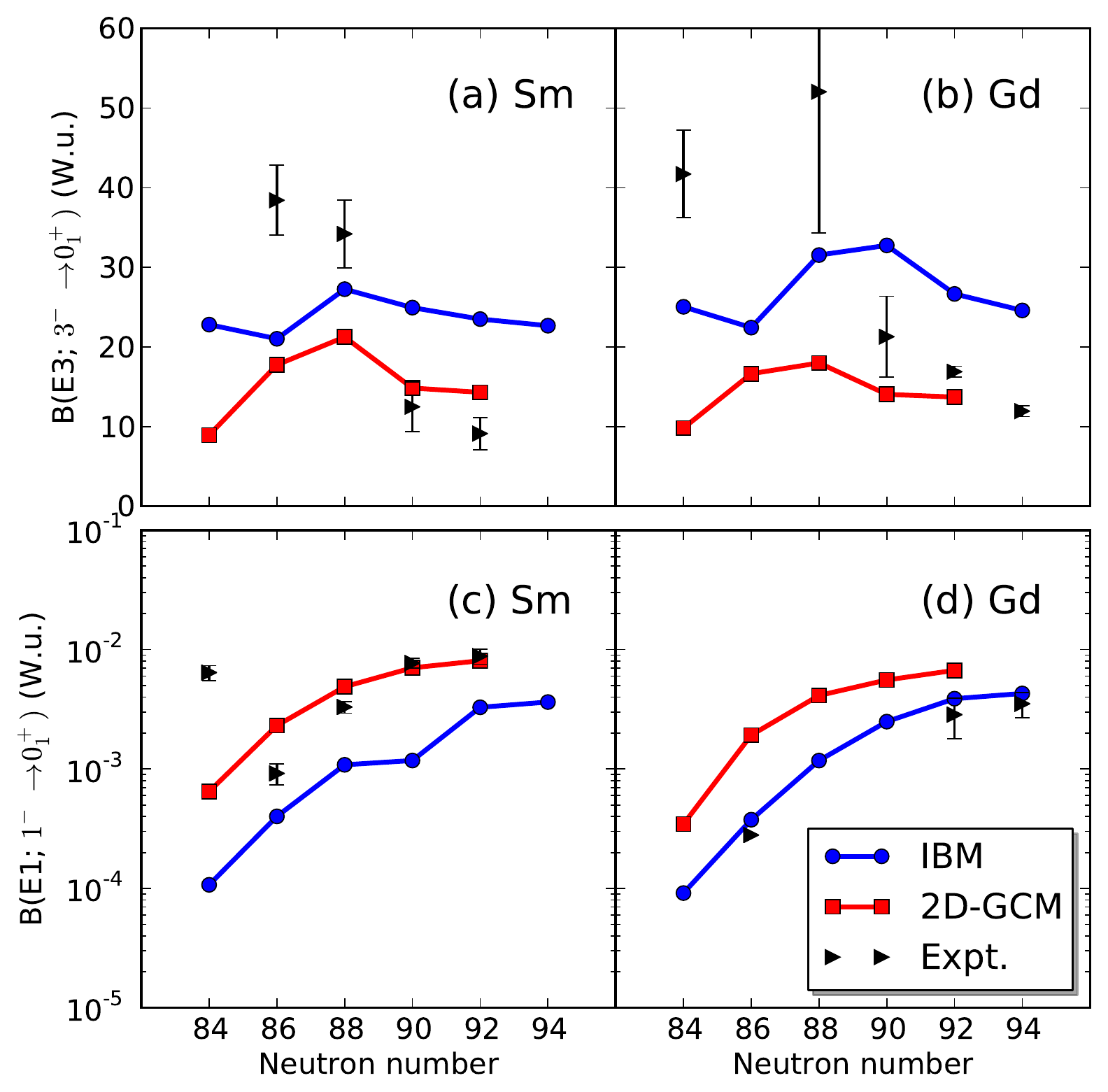}
\end{tabular}
\caption{
(Color online) Theoretical and experimental transition probabilities 
$B(E3;3^-_1\rightarrow 0^+_1)$ and $B(E1;1^-_1\rightarrow  0^+_1)$ for  
$^{146-156}$Sm and $^{148-158}$Gd. The experimental data are taken from
\cite{metzger76,data,kibedi02,pitz89}.} 
\label{fig:trans}
\end{center}
\end{figure*}

The reduced transition probabilities $B(E3;3^-_1\rightarrow 0^+_1)$ and 
$B(E1;1^-_1\rightarrow 0^+_1)$ are compared in Fig.~\ref{fig:trans} 
with the experimental data \cite{metzger76,data,kibedi02,pitz89}. For 
both isotopic chains, the predicted E3 transition rates [panels (a) and 
(b)] exhibit a weak dependence on the neutron number with a maximum at 
$N=88-90$. The down-sloping tendency in the theoretical (IBM) E3 values observed in the  
heavier isotopes is consistent with the experiment though a smoother 
change with neutron number is found for Sm isotopes. On the other hand, 
the E1 transition rates [panels (c) and (d)]  increase with increasing 
neutron number which agrees quite well with the experiment, exception 
made of $^{146}$Sm. The overall  trend also agrees well  with the one 
found in previous IBM \cite{nom14} and GCM \cite{rayner12} 
calculations. 
Note that the discrepancy of the IBM rates with the experimental ones
are partly  a consequence of the 
particular choice of the IBM effective charges. No effective charges 
are needed within the GCM framework  \cite{rayner12} as all the nucleons 
are considered in the wave functions.

%.......................................................................
\subsection{Spectroscopy of the nuclei $^{150}$Sm  and $^{158}$Gd}
\label{sec:spec}
%....................................................................... 

\begin{figure*}[ctb!]
\begin{center}
\begin{tabular}{cc}
\includegraphics[width=.8\linewidth]{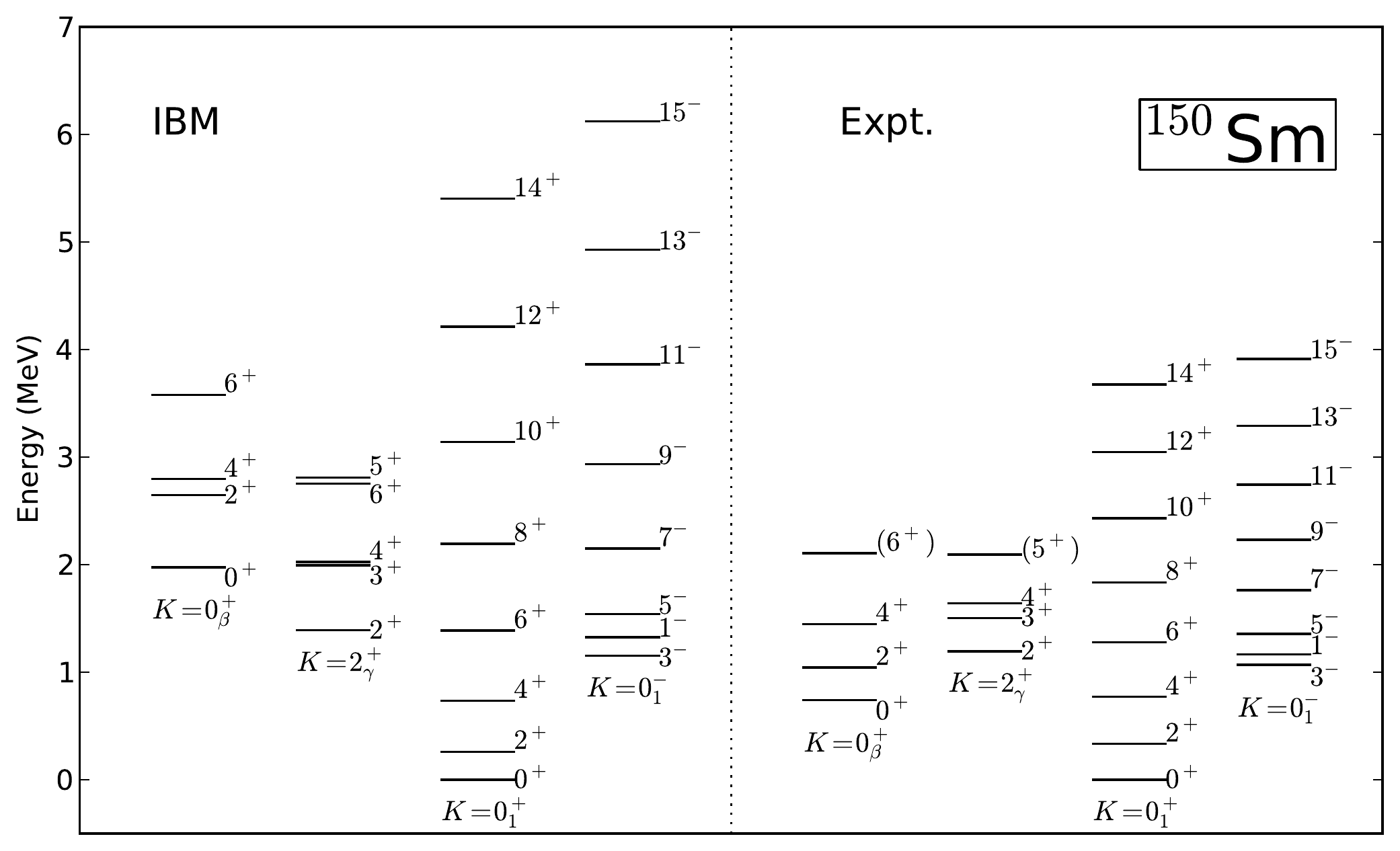}
\end{tabular}
\caption{
Comparison of the low-energy  spectrum predicted within 
the IBM framework for the nucleus $^{150}$Sm with the available 
experimental excitation energies \cite{data}. 
}
\label{fig:sm150}
\end{center}
\end{figure*}

\begin{table}[cb!]
\caption{\label{tab:sm150e2} 
Theoretical and experimental \cite{data} $B(\textnormal{E}2)$ transitions 
for $^{150}$Sm (in Weisskopf units). For details, see the main text.}
\begin{center}
\begin{tabular}{cccc}
\hline\hline
\textrm{$J_{i}^{\pi}$} &
\textrm{$J_{f}^{\pi}$} &
\textrm{$B(E2)_{\textnormal{theor.}}$} &
\textrm{$B(E2)_{\textnormal{expt.}}$} \\
%\textrm{$B(E\lambda; J^{\pi}_{i}\rightarrow J^{\pi}_{f})$ (W.u.)} \\
\hline
$2^+_1$ & $0^+_1$ & 79 & 57.1(13) \\
$4^+_1$ & $2^+_1$ & 112 & 110(17) \\
$6^+_1$ & $4^+_1$ &  120 & 1.5$\times 10^{+2}$(5)\\
$8^+_1$ & $6^+_1$ & 117 & 1.7$\times 10^{+2}$(9) \\
%$10^+_1$ & $8^+_1$ & - & 106 \\
%$12^+_1$ & $10^+_1$ & - & 91 \\
$0^+_\beta$ & $2^+_1$ &  10 & 53(5) \\
$2^+_\beta$ & $0^+_1$ & 1.78 & 0.81$^{+26}_{-21}$ \\
        & $0^+_\beta$ &  24 & 1.1$\times {10^2}^{+4}_{-3}$ \\
        & $2^+_1$ & 0.038 & -\\
        & $4^+_1$ &  3.92 & - \\
$2^+_\gamma$ & $0^+_1$ &  1.18 & 2.1(15) \\
        & $0^+_\beta$ &  10.2 & 9.1(24) \\
        & $2^+_1$ &  21 & - \\
        & $2^+_\beta$ &  4.09 & - \\
        & $4^+_1$ &  0.064 & 7(3) \\
$3^+_1$ & $2^+_\beta$ & 11.1 & - \\
$4^+_\beta$ & $2^+_1$ & 0.86 & - \\
        & $2^+_\beta$ & 11.2 & 1.9$\times 10^{+2}$(9) \\
        & $2^+_\gamma$ & 0.039 & 42(20) \\
        & $3^+_1$ & 0.34 & - \\
$4^+_\gamma$ & $2^+_1$ & 0.14 & 1.4(7) \\
        & $2^+_\beta$ &  2.1 & 4.1(21) \\
        & $2^+_\gamma$ &  42 & - \\
$1^-_1$ & $3^-_{K=0^-}$ &  109 & - \\
$5^-_{K=0^-}$ & $3^-_{K=0^-}$ &  70 & - \\
$7^-_{K=0^-}$ & $5^-_{K=0^-}$ &  84 & - \\
\hline\hline
\end{tabular}
\end{center}
\end{table}

% ----------------------------------------------------------------------

\begin{table}[cb!]
\caption{\label{tab:sm150e1} 
Same as in Table~\ref{tab:sm150e2}, but for the E1 transitions 
(in $10^{-3}$ W.u.).}
\begin{center}
\begin{tabular}{cccc}
\hline\hline
\textrm{$J_{i}^{\pi}$} &
\textrm{$J_{f}^{\pi}$} &
\textrm{$B(E1)_{\textnormal{theor.}}$} &
\textrm{$B(E1)_{\textnormal{expt.}}$} \\
\hline
$1^-_{{K=0^-}}$ & $0^+_1$ & 1.1 & 1.4$^{+7}_{-5}$ \\
        & $2^+_1$ & 0.13 &
		 2.9$^{+14}_{-10}$ \\
$3^-_{K=0^-}$ & $2^+_1$ & 2.5 & 5$^{+4}_{-3}$ \\
        & $4^+_1$ & 1.8$\times 10^{-4}$ &
		 5$^{+4}_{-3}$ \\
$4^+_\beta$ & $3^-_{K=0^-}$ & 2.6  & - \\
                    & $5^-_{K=0^-}$ & 0.24 & - \\
$4^+_\gamma$ & $3^-_{K=0^-}$ & 0.087 & 0.27(13) \\
                    & $5^-_{K=0^-}$ & 0.54 & 0.9(5) \\
$5^-_{K=0^-}$ & $4^+_1$ & 4.2 & - \\
$6^+_1$ & $5^-_{K=0^-}$ & 0.027 & - \\
$7^-_{K=0^-}$ & $6^+_1$ & 5.8 & - \\
$8^+_1$ & $7^-_{K=0^-}$ & 0.15 & - \\
%$8^+_2$ & $7^-_{K=0^-}$ & 7.7 & - \\
$9^-_{K=0^-}$ & $8^+_1$ & 7.3 & - \\
$10^+_1$ & $9^-_{K=0^-}$ & 0.46 & - \\
$11^-_{K=0^-}$ & $10^+_1$ & 9.0 & - \\
\hline\hline
\end{tabular}
\end{center}
\end{table}

% ----------------------------------------------------------------------

The low-lying spectrum of  $^{150}$Sm is compared in 
Fig.~\ref{fig:sm150} with the available experimental excitation 
energies \cite{data}. The band assignment has been made according to 
the dominant E2 transition sequence. The IBM  energies are generally 
more stretched than the experimental ones. Approximate alternating 
parity bands can be seen with the level ordering $7^-$, $8^+$, $9^-$, 
$10^+$, \ldots etc.

A noticeable deviation with respect to the experimental data is 
obtained for the $\beta$-vibrational band-head. In fact, the 
experimental excitation energy of this $0^+_2$ state is as small as the 
one for the $4^+_1$ state. However, in the calculations it is almost 
twice higher, suggesting a too  limited IBM model space. On the other 
hand, for the quasi-$\gamma$ band, with the $K^{\pi}=2^+$ built on 
the $2^{+}_2$ state, our  calculations predict the staggering 
($3^+_\gamma,4^+_\gamma$), ($5^+_\gamma,6^+_\gamma$), etc. This 
reflects the lack of triaxiality in the present study. The inclusion of 
mean-field triaxiality as well as the relevant  terms in the mapped IBM 
Hamiltonian could be useful to better describe the structure of the 
quasi-$\gamma$ band  \cite{Nom12tri}. Work along these lines is in 
progress and will be reported elsewhere.

The  E2 and E1 transition rates obtained for $^{150}$Sm are compared 
with the experimental  ones \cite{data} in Tables~\ref{tab:sm150e2} and 
\ref{tab:sm150e1}, respectively. Most of the predicted E2 values agree 
reasonably well with the experiment. Note that our calculations 
account for the $K=0^-$ band, built on the $3^-_1$ state, with strong 
E2 transitions. Nevertheless, large discrepancies are also found for 
some inter-band transitions. For example, the $0^+_\beta\rightarrow 
2^+_1$ strength is considerably underestimated.  Stronger inter-band E2 
transitions suggest a  significant mixing between different intrinsic 
configurations. Indeed, a recent experiment has suggested a complex 
shape coexistence in $^{152}$Sm \cite{garrett09}. Within this context, 
an IBM model space larger than the one considered in the present study 
may be required. A configuration mixing associated with  intruder 
states \cite{Nom12sc} could also be introduced to better describe a 
transitional nucleus like $^{150}$Sm. Another alternative could be the 
inclusion of triaxiality to better constrain the form of the  IBM 
Hamiltonian. Furthermore, the value $B(E2; 4_\beta^+ \rightarrow 
2^+_\gamma)=0.039$ W.u. is too small as compared with the experimental one 
[42(20) W.u]. A possible reason may be that the  $0^+_\beta$ states as 
well as the ones built on it might not be well described by the present 
calculations.

The calculated $B(E1)$ values  in Table~\ref{tab:sm150e1} reveal 
rather strong transitions (starting around the $J>5^-$) from the states 
of odd-$J$ negative-parity $K=0^-_1$ to those of the even-$(J-1)$ 
positive-parity ground-state bands. This fact, as well as the 
increasing $B(E1;J^-_{K=0^-}\rightarrow (J-1)^+_{K=0^+_1})$ value, as a 
function of $J$, signals the existence of an alternating parity band in 
$^{150}$Sm. Nevertheless, we do not consider the $B(E1)$ value obtained 
in the present calculation to be conclusive, mainly because of the lack 
of the $p$-boson effect in our framework. Indeed, as already  pointed 
out in previous phenomenological \cite{zamfir01} and microscopic 
\cite{taka86} studies on octupole-deformed nuclei, the description of 
these E1 transitions in the IBM framework could be improved by 
explicitly including the $p$ boson in the model space or by extending 
the form of the E1 operator so as to absorb the $p$-boson effect in the 
$sdf$ space. 

% ----------------------------------------------------------------------

\begin{figure*}[ctb!]
\begin{center}
\begin{tabular}{cc}
\includegraphics[width=.8\linewidth]{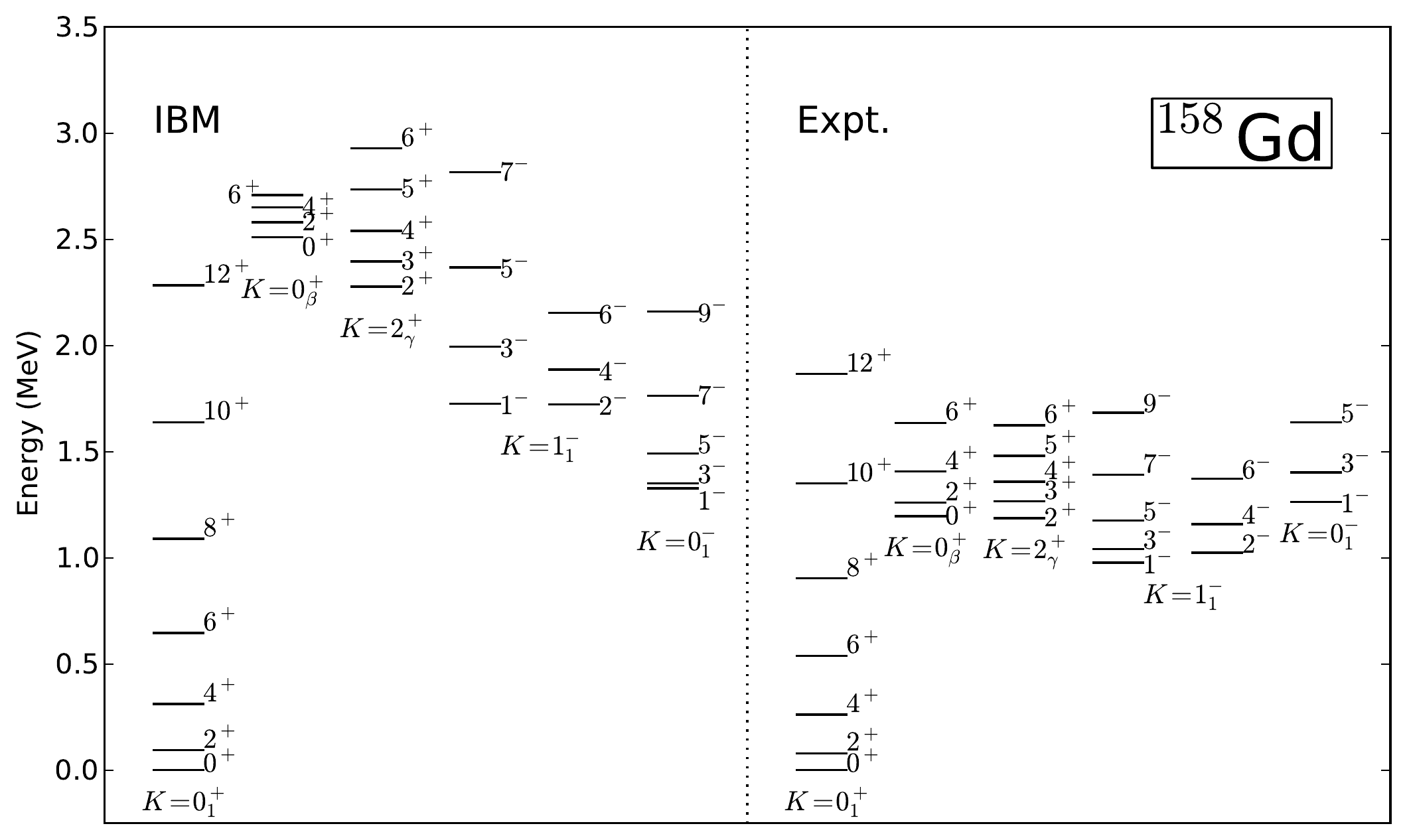}
\end{tabular}
\caption{
Same as in Fig.~\ref{fig:sm150} but for the nucleus 
$^{158}$Gd.}
\label{fig:gd158}
\end{center}
\end{figure*}

% ----------------------------------------------------------------------

% ----------------------------------------------------------------------

\begin{table}[cb!]
\caption{\label{tab:gd158e2} Same as in Table~\ref{tab:sm150e2} but for 
the nucleus $^{158}$Gd.}
\begin{center}
\begin{tabular}{cccc}
\hline\hline
\textrm{$J_{i}^{\pi}$} &
\textrm{$J_{f}^{\pi}$} &
\textrm{$B(E2)_{\textnormal{theor.}}$} &
\textrm{$B(E2)_{\textnormal{expt.}}$} \\
%\textrm{$B(E\lambda; J^{\pi}_{i}\rightarrow J^{\pi}_{f})$ (W.u.)} \\
\hline
$2^+_1$ & $0^+_1$ & 170 & 198(6) \\
$4^+_1$ & $2^+_1$ & 241 & 289(5) \\
$6^+_1$ & $4^+_1$ & 260 & - \\
$8^+_1$ & $6^+_1$ & 264 & 3.3$\times 10^{+2}$(3) \\
%$10^+_1$ & $8^+_1$ & - & 106 \\
%$12^+_1$ & $10^+_1$ & - & 91 \\
$2^+_\gamma$ & $0^+_1$ & 3.9  & 3.4(3) \\
        & $2^+_1$ & 7.7  & 6.0(7) \\
        & $4^+_1$ & 0.59  & 0.27(4) \\
$2^+_\beta$ & $0^+_1$ & 0.14  & 0.31(4) \\
        & $2^+_1$ & 0.088  & 1.39(15) \\
$4^+_\beta$ & $2^+_\gamma$ & 0.88  & 12.8 \\
        & $2^+_\beta$ & 93  & 455 \\
$3^-_{K=1^-}$ & $1^-_{K=1^-}$ & 105  & - \\
$3^-_{K=0^-}$ & $1^-_{K=0^-}$ & 146  & $>1.6\times 10^{+3}$ \\
$4^-_{K=1^-}$ & $3^-_{K=1^-}$ &  41  & 781(14) \\
$5^-_{K=1^-}$ & $3^-_{K=1^-}$ & 189  & 369(6) \\
$5^-_{K=0^-}$ & $3^-_{K=0^-}$ & 159  & - \\
$4^-_{K=1^-}$ & $2^-_{K=1^-}$ & 142  & 2.09$\times 10^{+3}$(3) \\
\hline\hline
\end{tabular}
\end{center}
\end{table}

% ----------------------------------------------------------------------

% ----------------------------------------------------------------------

\begin{table}[cb!]
\caption{\label{tab:gd158e1} Same as in Table~\ref{tab:sm150e2} but for 
the E1 transitions (in units of $10^{-3}$ W.u.) in $^{158}$Gd.}
\begin{center}
\begin{tabular}{cccc}
\hline\hline
\textrm{$J_{i}^{\pi}$} &
\textrm{$J_{f}^{\pi}$} &
\textrm{$B(E1)_{\textnormal{theor.}}$} &
\textrm{$B(E1)_{\textnormal{expt.}}$} \\
\hline
$1^-_{K=1^-}$ & $0^+_1$ & 1.5  & 0.098443(4) \\
        & $2^+_1$ & 2.8  & 0.096515(6) \\
$1^-_{K=0^-}$ & $0^+_1$ & 4.3  & 3.5(12) \\
        & $2^+_1$ & 2.3  & 6.4(21) \\
$3^-_{K=1^-}$ & $2^+_1$ & 0.35  & 0.33(10) \\
        & $4^+_1$ & 3.4  & 0.29(8) \\
$3^-_{K=0^-}$ & $2^+_1$ & 6.8  & $>$ 1.1 \\
        & $4^+_1$ & 0.74  & $>$ 1.5 \\
$2^-_{K=1^-}$ & $2^+_1$ & 4.5  & $<$ 0.078 \\
$4^-_{K=1^-}$ & $4^+_1$ & 4.5  & 0.090628(4) \\
%$5^-_1$ & $4^+_1$ &   & - \\
%$7^-_1$ & $6^+_1$ &   & - \\
%$8^+_1$ & $7^-_1$ &   & - \\
%$9^-_1$ & $8^+_1$ &   & \\
%$10^+_1$ & $9^-_1$ & & - \\
\hline\hline
\end{tabular}
\end{center}
\end{table}

% ----------------------------------------------------------------------

The low-lying spectrum of $^{158}$Gd, shown in Fig.~\ref{fig:gd158}, 
exhibits an overall agreement with the available experimental data for 
the lowest-lying positive- and negative-parity bands. The $1^-_1$ state 
of the lowest negative-parity band is assigned as the band-head of the 
$K^{\pi}=0^-_1$ and the $K^{\pi}=1^-_1$ bands in the present 
calculation and in the NNDC compilation \cite{data}, respectively.

In our calculations, the two lowest-lying positive-parity, 
$K^{\pi}=0^+_1$ and $2^+_\gamma$, bands are comprised of states with 
$n_f\approx 0.02$ and $0.06 \leqslant n_f\leqslant 0.09$, respectively, 
suggesting that they are almost pure  positive-parity bands. On the 
other hand, the states in the band built on the $0^+_2$ ($0^+_\beta$) 
state are of two-$f$ boson (equivalently double octupole phonon) nature 
with $\langle\hat n_f\rangle\approx 2$.  The side-band energies, 
especially for those states in the positive-parity $\beta$-vibrational 
and quasi-$\gamma$ bands, are overestimated considerably, for similar 
reasons as in the $^{150}$Sm case.

In Tables.~\ref{tab:gd158e2} and \ref{tab:gd158e1}, we have compared 
some relevant E2 and E1 transition rates with the experimental ones 
\cite{data}. Many of the calculated E2 transition rates agree well with 
the data. Again a noticeable deviation is observed for the 
$4^+_\beta\rightarrow 2^+_\gamma$ transitions, probably for the same 
reason as in  the $^{150}$Sm case. 
We note that the lifetime of the experimental $3^{-}_{K=1^-}$ state
adapted in \cite{data} has nearly 25\% of uncertainty, and that, for
that reason, the
error bars for the reduced E2 transitions $4^{-}_{K=1^-}\rightarrow
3^{-}_{K=1^-}$ and $5^{-}_{K=1^-}\rightarrow 3^{-}_{K=1^-}$ shown in
Ref.~\cite{data}, as well as in Table~\ref{tab:gd158e2}, could be
corrected. 
From Table~\ref{tab:gd158e1} one 
concludes that our model gives a reasonable description of the E1 
transitions associated to  states in the $K^\pi=0^-_1$ band whose 
energies are described rather nicely as well (see 
Fig.~\ref{fig:gd158}). However, our model does not account for the E1 
transitions associated to the  $K^\pi=1^-_1$ band.

% ----------------------------------------------------------------------

\subsection{Excited $0^+$ states}
\label{excited-zeros}

% ----------------------------------------------------------------------

\begin{figure*}[ctb!]
\begin{center}
\begin{tabular}{cc}
\includegraphics[width=14cm]{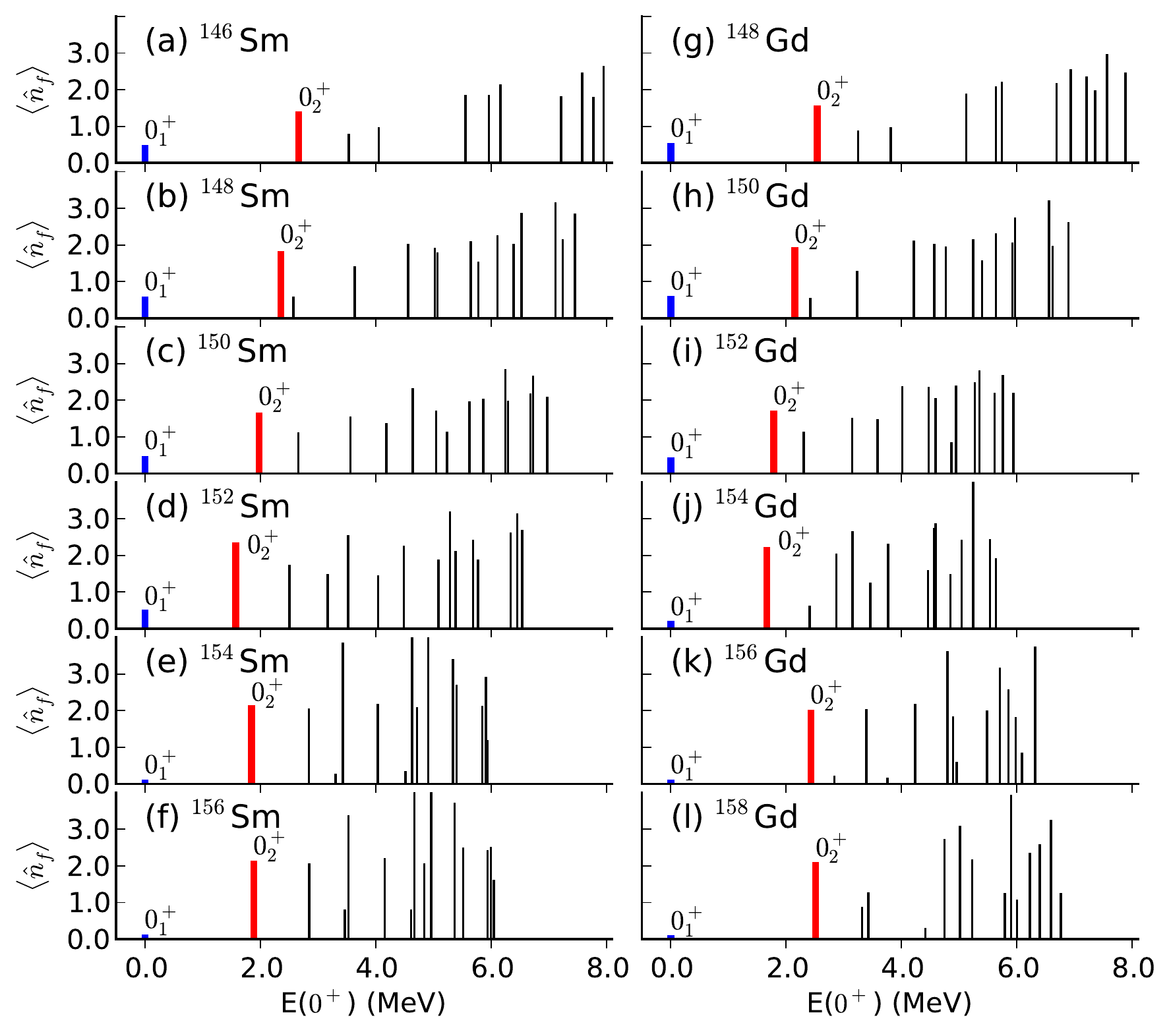}
\end{tabular}
\caption{(Color online) Energy distribution of the theoretical
 lowest fifteen $0^+$ states 
 and expectation value of the $f$-boson number 
 operator $\langle\hat n_f\rangle$
 for the considered Sm and Gd isotopes. Note that, concerning $^{146}$Sm [panel
 (a)] and 
 $^{148}$Gd [panel (g)], the $0^+$ states with an energy higher than 8 MeV are not shown.}
\label{fig:nf}
\end{center}
\end{figure*}

% ----------------------------------------------------------------------

Experimentally, many excited $0^+$ states have been identified in the
low-energy excitation spectrum of $^{158}$Gd. The previous 
phenomenological calculation within the $spdf$ IBM model 
\cite{zamfir02} showed that such a large number of excited $0^+$ states 
at relatively low energy can be described if the octupole degrees of 
freedom is taken into account, and many of the $0^+$ states have been 
attributed to the coupling of two octupole phonons. 
Meanwhile, the emergence of a large number of low-energy excited $0^+$ states can be a
good signature of a quantum phase transition \cite{meyer06}.

To address the nature of the $0^+$ states resulting from the mapped 
$sdf$ IBM Hamiltonian, we show in Fig.~\ref{fig:nf} the energy
distribution (or level scheme) of the lowest fifteen $0^+$ states and the corresponding average values 
of the $f$-boson number operator $\langle\hat n_f\rangle$ for the 
$^{146-156}$Sm [from panel (a) to panel (f)] and  
$^{148-158}$Gd [from panel (g) to panel (l)] nuclei. In the
$^{146}$Sm [panel (a)] and $^{148}$Gd [panel (g)] cases those states with an 
energy higher than 8 MeV are not shown. 
In all the nuclei,  the $0^+$ ground-state 
is predominantly composed of positive-parity ($s$ and $d$) bosons as 
$\langle\hat n_f\rangle<0.5$. 
In both isotopic chains, for many of the 
nuclei with $N\geqslant 90$, $\langle\hat n_f\rangle\approx 2$ for the 
$0^+_2$ state, suggesting its double-octupole phonon nature. 
Moreover, many other $0^+$ states are also formed by the
coupling of positive- and negative-parity (octupole) bosons. 
For both Sm and Gd chains, the $0^+$ states become more populated in
lower-energy region for the heavier isotopes, where the 
quadrupole-octupole coupling becomes more enhanced. 
Particularly in the Gd isotopes, the level scheme for the $0^+$ states becomes most compressed around 
$^{152}$Gd [panel (i)] or $^{154}$Gd [panel (j)], where the
corresponding potential energy surface is noticeably soft both in $\beta_2$ and
$\beta_3$ deformations [see, Figs.~\ref{fig:hfb_pes}(i,j)]. 
%However, the $\langle\hat n_f\rangle$ values for the $0^+$ states do 
%not show a clear tendency suggesting an important role of the 
%quadrupole-octupole coupling. In addition, those values in the Sm 
%isotopes look very different from the ones for the Gd isotopes. 

% ----------------------------------------------------------------------
\subsection{Correlation energy}
\label{sec:corr}
% ----------------------------------------------------------------------

% ----------------------------------------------------------------------

\begin{figure*}[ctb!]
\begin{center}
\includegraphics[width=0.65\linewidth]{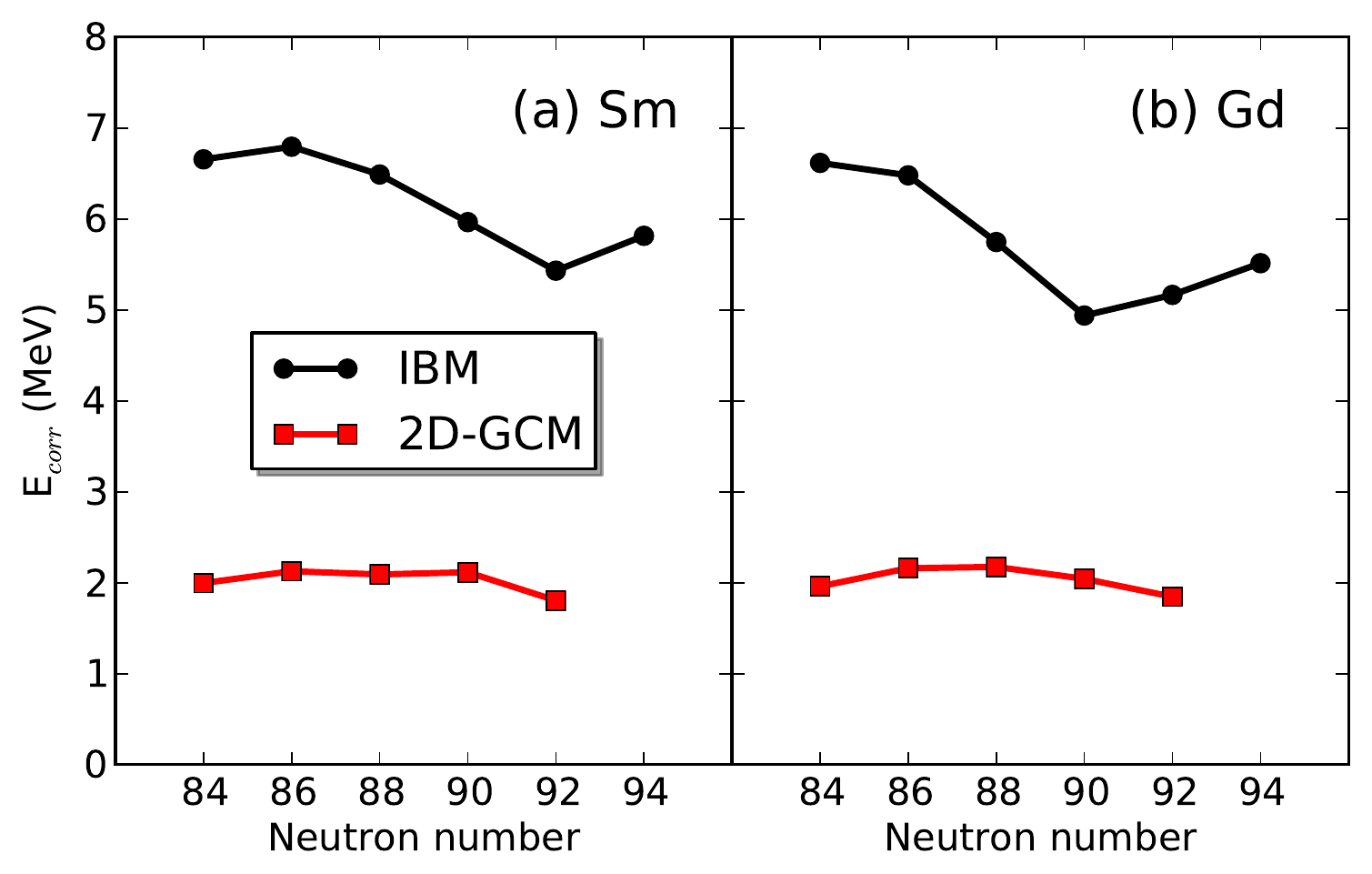}
\caption{(Color online) 
The correlation energies obtained from the IBM and the two-dimensional
 (2D) GCM for $^{146-156}$Sm and $^{148-158}$Gd isotopes. 
}
\label{fig:corr}
\end{center}
\end{figure*}

% ----------------------------------------------------------------------

In this section, we discuss the correlation energies defined as
\cite{Nom10,rayner12}
\begin{eqnarray}
 E_{\textnormal{Corr}}=E^{g.s.}_{\textnormal{HFB}}-E(0^+_1). 
\end{eqnarray}
where $E^{g.s.}_{\textnormal{HFB}}$ represents the HFB ground-state 
energy and $E(0^+_1)$ the one for the  $0^+_1$ state. 
The IBM correlation energies are depicted in 
Fig.~\ref{fig:corr} together with the ones obtained in previous 
two-dimensional Gogny-D1M GCM calculations. Results are shown in panel 
(a) for $^{146-156}$Sm and in panel (b) for $^{148-158}$Gd. Though the 
correlation energies are different in both approaches, the largest 
values of  $E_{\textnormal{Corr}}$ are obtained for the lighter nuclei 
with $N\leqslant 88$ which are rather soft in the $\beta_2$ and $\beta_3$ 
degrees of freedom. This confirms that  correlations beyond the 
mean-field approach can become significant in soft nuclear systems. 
%In both the IBM and GCM frameworks $E_{\textnormal{Corr}}$ decreases for 
%heavier isotopes.

% ----------------------------------------------------------------------

\section{Summary\label{sec:summary}}
%\ref{sec:summary}

% ----------------------------------------------------------------------

In summary, we have carried out spectroscopic calculations aimed to 
describe  the quadrupole and octupole collective states in Sm and Gd 
isotopes. Our starting point was a set of $Q_{20}-Q_{30}$ constrained 
HFB calculations, with the D1M parametrization of the Gogny effective 
interaction, used to produce a potential energy surface. This potential 
energy surface is then used to obtain the parameters of an IBM 
Hamiltonian including $s$, $d$ and $f$ bosons. Spectral properties of 
both positive- and negative-parity states associated to the reflection 
symmetric and asymmetric shapes, respectively, are obtained after 
diagonalization of the IBM Hamiltonian. The parameters of the IBM 
Hamiltonian are determined by mapping the Gogny-HFB mean-field energy 
surface onto the corresponding energy expectation value of the boson 
condensate state. 

The systematics of the  energy spectra and transition rates, associated 
to both positive- and negative-parity yrast states, points to the onset 
of notable octupole correlation around $N\approx 88$, characterized by 
the $\beta_3$-soft energy surfaces (Fig.~\ref{fig:hfb_pes}), and the 
corresponding negative-parity band lowering in energy with respect to 
the positive-parity ground-state band (Fig.~\ref{fig:neg}). From 
$N\geqslant 90$ on, the potential energy surface no longer exhibits 
$\beta_3$ softness, and the corresponding negative-parity band is 
pushed up in energy with respect to the ground-state band.  The 
mean-field $\beta_2\beta_3$ energy surface (Fig.~\ref{fig:hfb_pes}), 
the derived parameters in the $sdf$ Hamiltonian (Fig.~\ref{fig:para}), 
the resultant energy levels (Figs.~\ref{fig:pos} and \ref{fig:neg}) and 
transition rates (Fig.~\ref{fig:trans}) correlate very well with each 
other in systematics with the number of valence nucleons. In addition, 
the spectroscopic properties resulting from the model turn out to be 
generally in a reasonable agreement with the systematics of the 
available experimental data, and also to be consistent with the 
previous GCM calculation (Figs.~\ref{fig:ibmgcm} and \ref{fig:trans}) 
starting from the common Gogny parametrization  D1M \cite{rayner12}. 

On the other hand, an in-depth analysis of the energy spectra and the 
E2 and E1 transition rates in the two characteristic cases, $\beta_2$- 
and $\beta_3$-soft nucleus $^{150}$Sm and strongly $\beta_2$ deformed 
nucleus $^{158}$Gd, has revealed that an improvement of the model is 
required so as to give a better description not only of the yrast 
states but also of the non-yrast states. For example, our model in its 
current version is not able to describe well the band-head of side 
bands, particularly that of the $\beta$-vibrational ($K^{\pi}=0^+_2$) 
band (Figs.~\ref{fig:sm150} and \ref{fig:gd158}). A possible reason 
could be that the model space used for the present work might be rather 
limited to handle such a complex nuclear structure. This would require 
the extension of our model space to include configuration mixing 
specific to the intruder state and/or to introduce triaxial degrees of 
freedom. In addition, the model has failed in reproducing some of the 
E1 properties, especially for those associated to the states in 
non-yrast negative-parity band (Fig.~\ref{fig:gd158}). Several 
solutions have been proposed that could help to fix the problem: 
extension of the E1 operator to include higher-order terms; explicit 
inclusion of $p$ boson in the model space. Improving the description of 
these properties will be a topic of future study. 
Significance of the $p$-boson effect in the E1 excitation observed in
rare-earth nuclei has been addressed in \cite{spieker15},
though in the different context of $\alpha$ clustering. 
 
We have also analyzed the wave function content of some lower-lying 
excited $0^+$ states for the considering nuclei, and found that in many 
of the nuclei considered, the $0^+_2$ states can be the consequence of 
the coupling of two-octupole phonons. This could be a  possible 
explanation for the large number of low-energy excited $0^+$ states 
found in rare-earth nuclei. 

\begin{acknowledgements}
K. N. acknowledges the support by the Marie Curie Actions grant within
the Seventh Framework Program of the European Commission under Grant
No. PIEF-GA-2012-327398. The work of LMR is  supported in part by 
Spanish  MINECO grants Nos. FPA2012-34694 and FIS2012-34479
and by the Consolider-Ingenio 2010 program MULTIDARK CSD2009-00064.
\end{acknowledgements}

\bibliography{refs}
\end{document}